\documentclass[]{aa}  

\usepackage{graphicx}
\usepackage{txfonts}
\usepackage{lipsum}
\usepackage{subcaption}         
\usepackage{hyperref}
\usepackage{siunitx}
\usepackage{natbib}
\usepackage{lscape}             
\usepackage{placeins}           

\usepackage{pifont}
\usepackage{tabularx}

\newcommand{\argmin}{\mathop{\mathrm{arg\:min}}}

\begin{document}

   \title{Enhancing Image Resolution of Solar Magnetograms: A Latent Diffusion Model Approach}

   \author{F. P. Ramunno
          \inst{1, 2}\thanks{Email address: francesco.ramunno@fhnw.ch}
          \and
          P. Massa\inst{1}
          \and
          V. Kinakh\inst{2}
          \and
          B. Panos\inst{1}
          \and
          A. Csillaghy\inst{1}
          \and
          S. Voloshynovskiy\inst{2}
          }

   \institute{Institute for Data Science, University of Applied Sciences North Western Switzerland (FHNW), 5210 Windisch, Switzerland
         \and
             Department of Computer Science, University of Geneva, 1211 Geneva, Switzerland
             }

   \date{}

 
  \abstract{The spatial properties of the solar magnetic field are crucial to decoding the physical processes in the solar interior and their interplanetary effects. However, observations from older instruments, such as the Michelson Doppler Imager (MDI), have limited spatial or temporal resolution, which hinders the ability to study small-scale solar features in detail. Super resolving these older datasets is essential for uniform analysis across different solar cycles, enabling better characterization of solar flares, active regions, and magnetic network dynamics. In this work, we introduce a novel diffusion model approach for Super-Resolution and we apply it to MDI magnetograms to match the higher-resolution capabilities of the Helioseismic and Magnetic Imager (HMI). By training a Latent Diffusion Model (LDM) with residuals on downscaled HMI data and fine-tuning it with paired MDI/HMI data, we can enhance the resolution of MDI observations from 2"/pixel to 0.5"/pixel. We evaluate the quality of the reconstructed images by means of classical metrics (e.g., PSNR, SSIM, FID and LPIPS) and we check if physical properties, such as the unsigned magnetic flux or the size of an active region, are preserved. We compare our model with different variations of LDM and Denoising Diffusion Probabilistic models (DDPMs), but also with two deterministic architectures already used in the past for performing  the Super-Resolution task. Furthermore, we show with an analysis in the Fourier domain that the LDM with residuals can resolve features smaller than 2", and due to the probabilistic nature of the LDM, we can asses their reliability, in contrast with the deterministic models. Future studies aim to super-resolve the temporal scale of the solar MDI instrument so that we can also have a better overview of the dynamics of the old events.}  


   \maketitle

\section{Introduction}
\label{sec:introduction}
Understanding the spatial and temporal properties of the solar magnetic field is critical to decoding the physical processes within the solar interior, its atmosphere, and its impact on Earth \citep{Wiegelmann2014, Wei2021, Wang2023, Yadav_2023, GEORGOULIS2024}. 

The magnetic field is accurately measured in the solar photosphere, and several space missions have observed it with different spatial and temporal resolutions. The higher the spatial resolution, the better we can characterise the morphology of small-scale features, while the higher the temporal resolution the easier will be to understand the evolution of their physical processes \citep{Wiegelmann2014}. In addition, studying the solar photosphere in different solar cycles can be beneficial to truly understand the properties of the magnetic network \citep{Li2019}. However, despite the amount of data that covers different solar cycles, they come from different instruments with different spatial and temporal resolutions, and the older the instruments are, the lower their capabilities. Consequently, it is difficult to perform an overall analysis due to the lack of uniformity. For this reason, having the possibility to translate the data seen by an older telescope as a new telescope would have seen them can be beneficial for overcoming the hardware limitations of the past \citep{Liu2012, Virtanen2019, Kinakh2024}. 

In deep learning, Super-Resolution refers to enhancing the resolution of images by increasing their spatial dimensions \citep{SU2025110935}. The key objective is to predict missing high-frequency details that are not present in the low-resolution version. Super-Resolution has been widely analysed in the field of Computer Science in the last decade \citep{Saharia2021, Rombach2021, Pernias2023}, resulting in a vast amount applications in astronomy \citep{Kinakh2024, Jarolim2024} since it can be used not only to increase the spatial dimensions but to improve the image under various noise conditions \citep{Armstrong2021, Chaoui2024}.

\begin{figure*}
\centering
 \includegraphics[width=0.9\hsize]{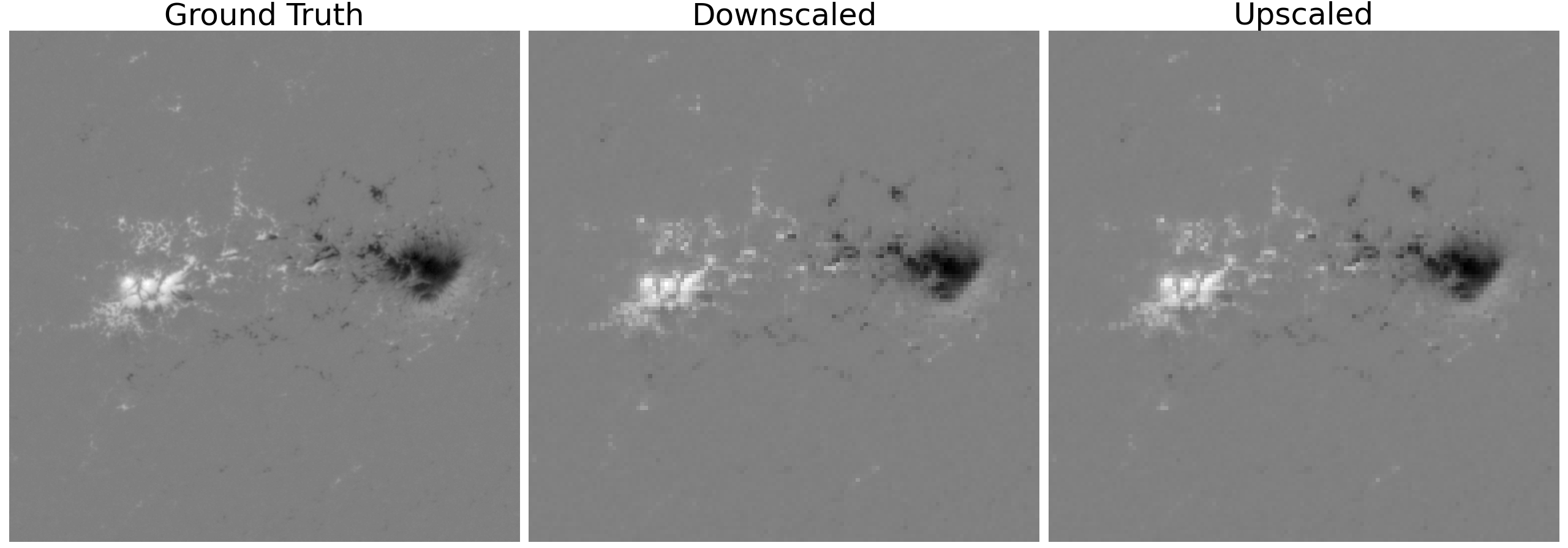}%
\caption{Example of a training pair composed of the ground truth image, its downscaled version, and the resulting uspcaled image which is obtained by replicating each pixel values of the downscaled image by 4. The latter image is provided as input to our model. We refer to Section \ref{sec:data_source} for more details.}
\label{fig:downscaled}
\end{figure*}

An excellent science case for applying Super-Resolution is related to two space-based instruments, the Michelson Doppler Imager (\citealp[MDI]{mdi}) on-board of the Solar and Heliospheric Observatory (SOHO) and the Helioseismic and Magnetic Imager (\citealp[HMI]{hmi}) on-board of the Solar Dynamics Observatory (SDO). The MDI/SOHO instrument launched in 1995 and operative up to April 2011 has observed the photosphere during Solar Cycle 23 with a spatial resolution of 2"/pixel and a temporal resolution of 96 minutes for the full disk Line of Sight (LoS) magnetograms. The HMI/SDO instrument launched in 2010 and still operating is observing the photosphere during Solar Cycle 24 and Solar Cycle 25 with a spatial resolution of 0.5"/pixel and a temporal resolution of 12 seconds for the full disk LoS magnetograms.
Various attempts were made to uniform the dataset provided by the two instruments \citep{Rahman2020, Xu2024, Jaramillo2024} in a deterministic behaviour by training the model with a pixel-wise loss. Unfortunately, the Super-Resolution problem is ill-posed, and each low-resolution (LR) image corresponds to infinite high-resolution (HR) images. Thus, training a model to minimize the mean squared error between predicted and target images for a set of examples, results in output images that represent an average prediction over the set of feasible predictions and therefore lack of fine-grained details \citep{bruna2016superresolutiondeepconvolutionalsufficient, ledig2017photorealisticsingleimagesuperresolution}. Moreover, there is no straightforward way to compute uncertainties on the output of deterministic models.

Denoising Diffusion Probabilistic Models (DDPMs) \citep{ho2020} have demonstrated high capabilities in Super-Resolution (SR) tasks \citep{Saharia2021} overcoming the Generative Adversarial Networks (GANs) \citep{Goodfellow2014} in terms of quality of the prediction and simplicity of the training task \citep{Dhariwal2021}. DDPMs are less prone to generate artefacts and it is possible to make use of their probabilistic nature to determine an uncertainty on the prediction, which is fundamental for scientific purposes \citep{Ramunno2024, Ramunno2024mag2mag}. Their stability arises from the iterative generation during inference \citep{Sun2024}. However, this rapidly increases computational demands with the image size \citep{Rombach2021}. Therefore, it has been demonstrated that it is possible to train DDPMs in a latent space of a pre-trained autoencoder \citep{Rombach2021} that reasonably represents the data. This approach leads to a reduction of the image size. Therefore, it allows implementing more complex network architectures with limited computational burden, and it permits the generation of images of the same size as the HMI/SDO telescope, 4096 $\times$ 4096 pixels with 0.5"/pixel as spatial resolution.

In this work, we train a latent diffusion model on downscaled HMI/SDO data, with a spatial resolution of 2"/pixel, to super-resolve them into the high-quality HMI/SDO data with a spatial resolution of 0.5"/pixel. Our novel method allows super-resolving features smaller than 2", a capability not found in other deterministic models. We also develop a technique to determine the reliability of these predicted features, making them more relevant from a physical point of view. This opens up the exciting possibility of super-resolving the data provided by MDI between 1995 and 2010 and studying the Solar Cycle 23 with the same resolution as HMI/SDO. Therefore, we now have the unique opportunity to study more eruptive events (in addition to those of Solar Cycle 24) with a higher spatial resolution.
In the future, we are interested in super-resolving the temporal resolution of MDI/SOHO, which would lead to a better understanding of the dynamic properties of the features on the photosphere.

This paper is organised as follows. In Section \ref{sec:data_source} we introduce the datasets used. In Section \ref{sec:background} we introduce the Super-Resolution problem and explain the latent diffusion model together with the palette technique for image-to-image translation and the two deterministic approaches considered. In Section \ref{sec:experiments} we analyse our setup and our experiments; we discuss the evaluation metrics in Section \ref{sec:metrics} and we discuss the results in Section \ref{sec:results}. In the end we apply the model to super-resolve MDI magnetograms and introduce a technique in the Fourier space to validate our results in Section \ref{sec:mdi_app} and we conclude in Section \ref{sec:conclusions}.

\section{Dataset}
\label{sec:data_source}

We consider data from two space-based instruments: the HMI/SDO \citep{hmi} and the MDI/SOHO \citep{mdi}. There is only a small time window in which HMI and MDI operated simultaneously from May 1 2010, to April 11 2011, with 4126 pairs of MDI and HMI data. To ensure more diversity in the training dataset and not only use data recorded in this short time range, we pre-train our model to a dataset made of only HMI data and then finetune it on the dataset shared among the two instruments. 
We construct a dataset of HMI images recorded between 2013 and 2019. For each image, we create its downscaled version, just averaging every 4 pixels and obtaining from a 4096 $\times$ 4096 pixel image a 1024 $\times$ 1024 pixel image. The downscaling procedure allows us to obtain a 2''/pixel spatial resolution, which reflects the spatial resolution of the MDI instrument. The final dataset consists of 43912 paired images of downscaled and actual HMI data. The image values are constrained to a range between -3000 and +3000 G, as the instrument's dynamic range limitations only become significant near 3000 G \citep{Hoeksema_2014}. Finally, the data are normalized to a range between -1 and 1.

\section{Background}
\label{sec:background}

\begin{figure*}
\centering
\subfloat[Training step\label{fig:training}]{%
 \includegraphics[width=1\linewidth]{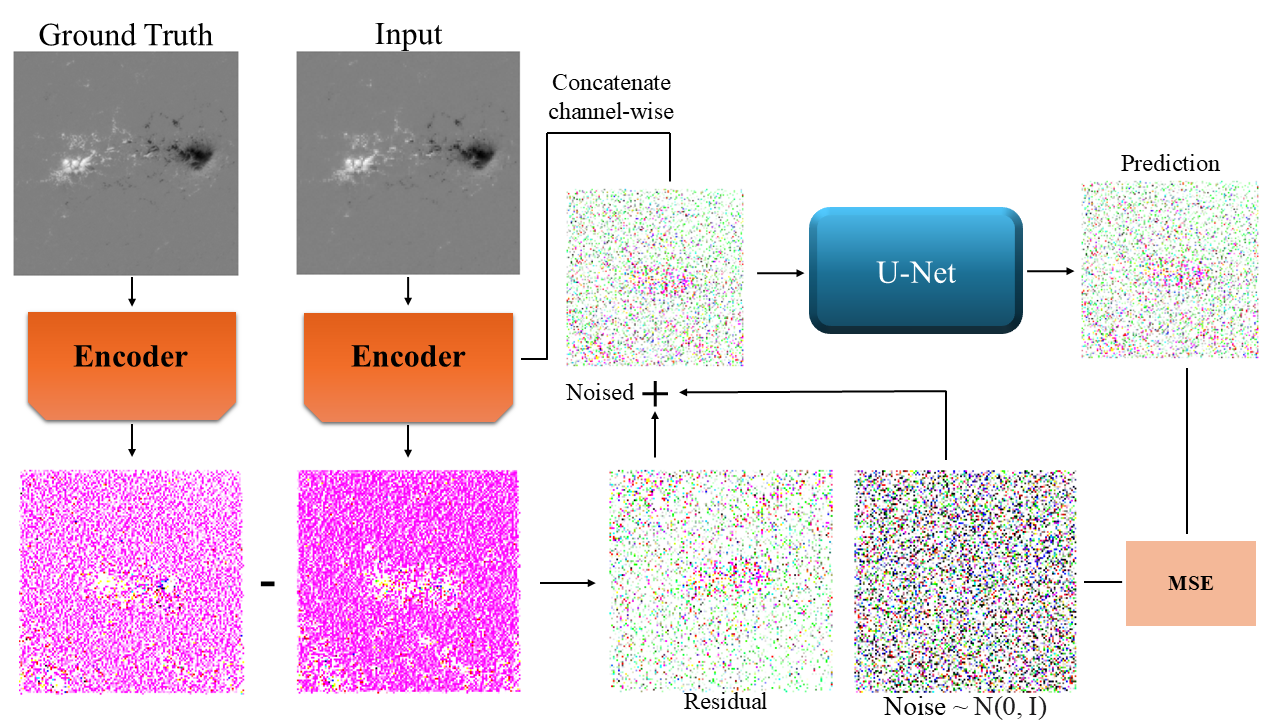}%
}\hspace{0.05\linewidth}
\subfloat[Inference step\label{fig:inference}]{%
 \includegraphics[width=1\linewidth]{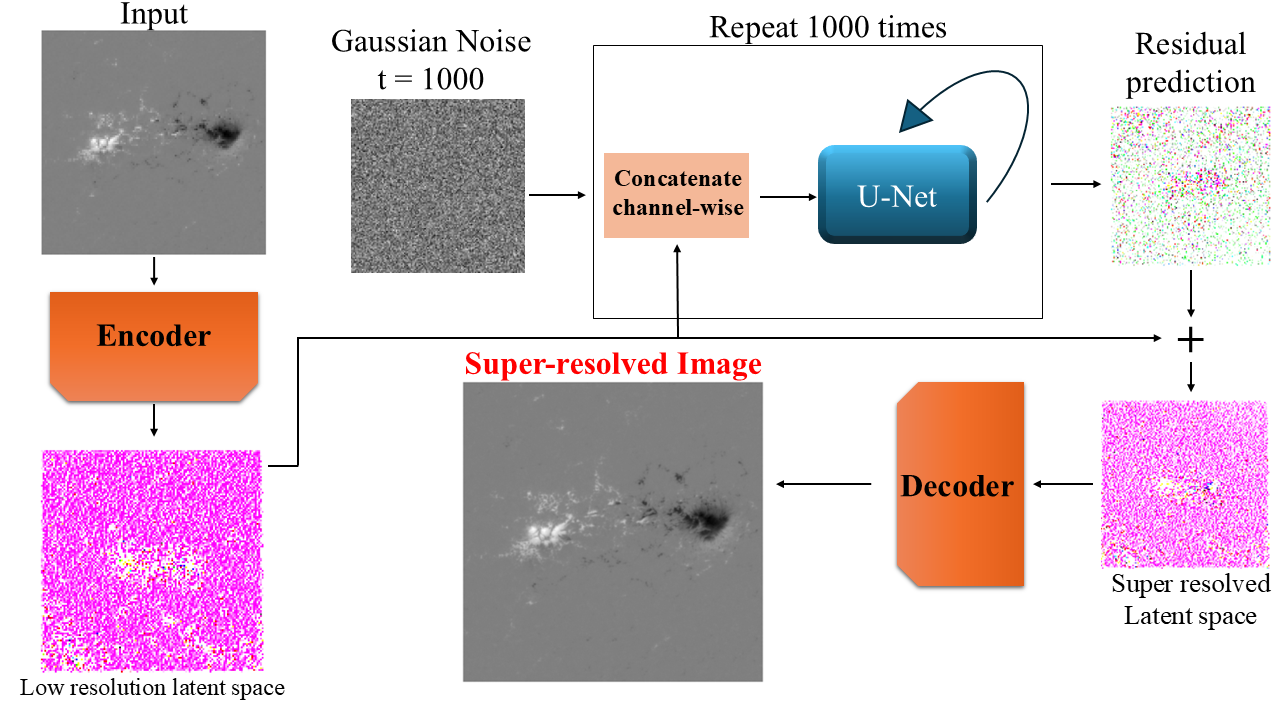}%
}
\caption{(a) The top image demonstrates the training process of the Latent Diffusion Residual Model (LDM) for image Super-Resolution. In this process, both the Input and Ground Truth images are passed through an Encoder; subsequently, the residual (difference) between them is calculated and injected with noise with a random magnitude determined by the timestep parameter t. This noisy latent representation is then processed by the LDM, which predicts the injected noise. Finally, Mean Squared Error (MSE) loss is computed between the predicted noise and the original injected noise.
(b) The bottom image illustrates the inference algorithm of the LDM for Super-Resolution. Here, the Input is encoded and concatenated channel-wise with Gaussian noise which corresponds at t=1000. The prediction process iterates 1,000 times, refining the residual prediction, which is ultimately added back to the encoded input image. The result is then decoded to produce the super-resolved image.}
\label{fig:concept_and_architecture}
\end{figure*}

\begin{figure*}
\centering
 \includegraphics[width=\hsize]{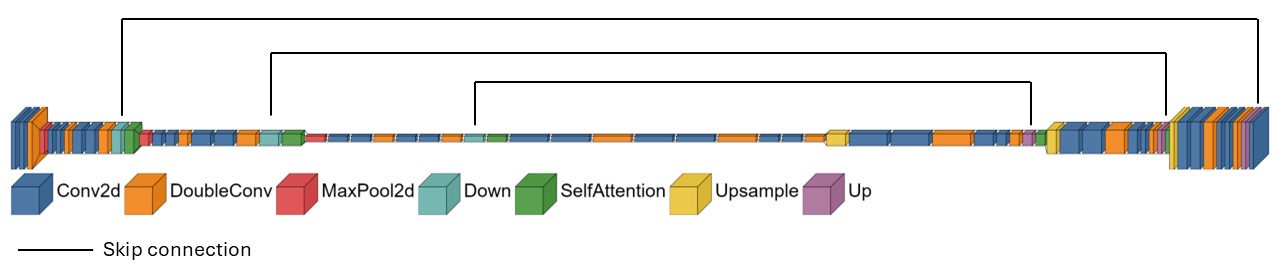}%
\caption{Detailed visual breakdown of the U-Net network architecture, consisting of convolutional layers, max-pooling, downsampling, self-attention, and upsampling layers, which enhance the image resolution by leveraging both spatial and feature-based processing techniques.}
\label{fig:architecture}
\end{figure*}

This section briefly introduces the Super-Resolution problem \citep{SU2025110935} and analyses its issues. We explain the Palette approach \citep{Saharia2021} that we used to condition our model, and finally, we describe the functioning of the latent diffusion model based on the approach suggested in \cite{Rombach2021}.

\subsection{Super resolution}
\label{sec:sr}
The goal of image Super-Resolution \citep{SU2025110935} is to transform a Low-Resolution (LR) image into a High-Resolution (HR) image, recovering the missing high frequency details. 

Given a LR image $x \in \mathbb{R}^{\hat{w} \times \hat{h} \times c}$, where $\hat{w}$, $\hat{h}$ and $c$ are respectively the height, the width and the number of channels of the image, the goal is to generate the corresponding HR image $y \in \mathbb{R}^{w \times h \times c}$, where $\hat{w} < w$ and $\hat{h} < h$. The relationship is represented by a degradation mapping:
\begin{equation}
    x = \mathcal{D}(y; \Theta),
\end{equation}
where $\mathcal{D}: \mathbb{R}^{w \times h \times c} \rightarrow \mathbb{R}^{\hat{w} \times \hat{h} \times c}$ is generally unknown, and $\Theta$ is the set of degradation parameters governing this mapping. The goal of a SR model is to determine the inverse mapping of $\mathcal{D}$ with a parametrized function $\hat{y} = f_{\theta}(x) = \mathbb{R}^{\hat{w} \times \hat{h} \times c} \rightarrow \mathbb{R}^{w \times h \times c}$, where $\theta$ indicates the array of parameters. The optimal parameter values are determined by solving:
\begin{equation}
    \theta^{*} = \argmin_{\theta}\mathcal{L}(f_{\theta}(x), y),
\end{equation}
where $\mathcal{L}$ represents the distance between the predicted HR image and the actual HR image.

\begin{figure*}
\centering
\subfloat[\label{fig:psnr}]{%
 \includegraphics[width=0.38\linewidth]{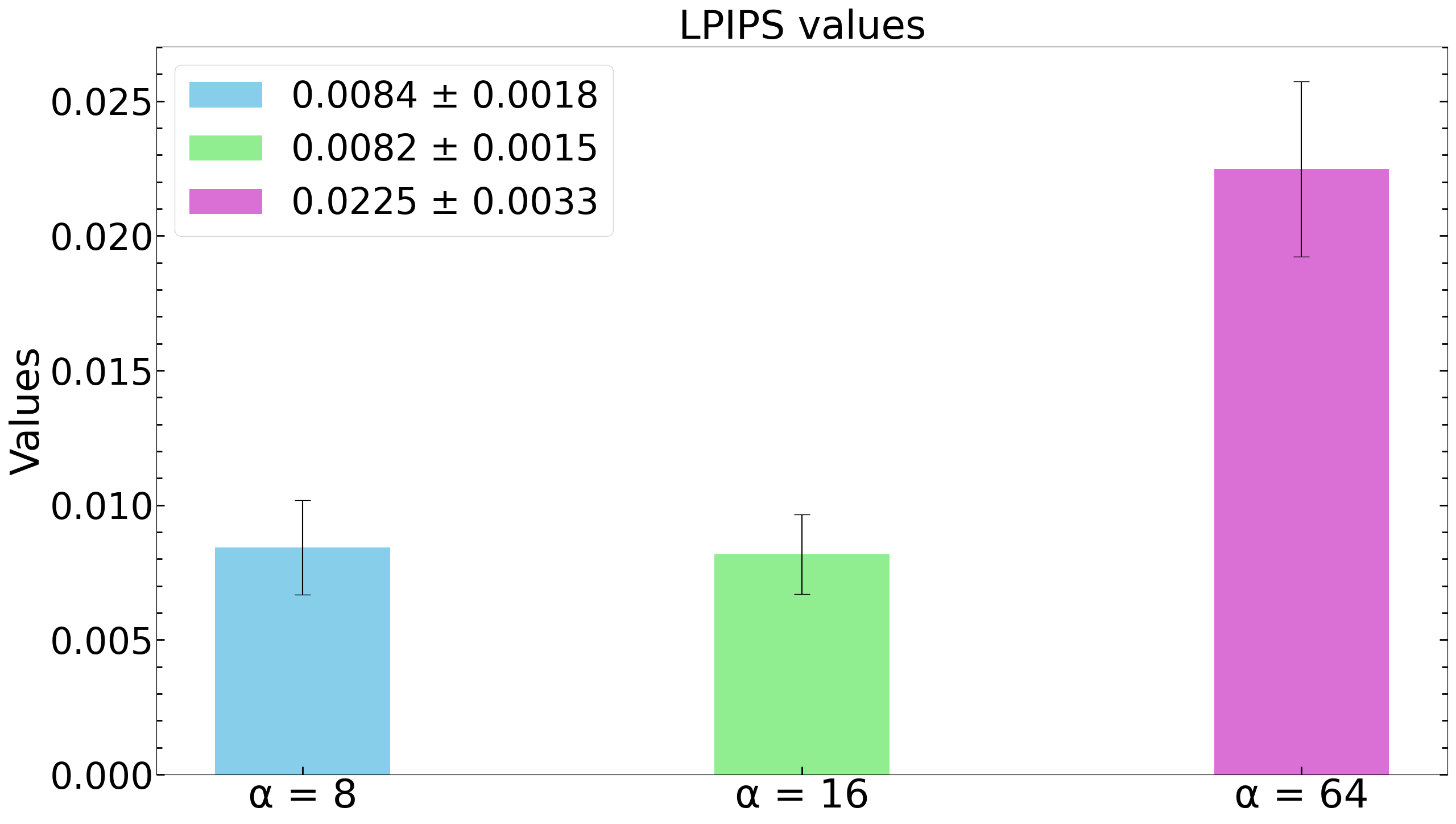}%
}\hspace{0.05\linewidth}
\subfloat[\label{fig:ssim}]{%
 \includegraphics[width=0.38\linewidth]{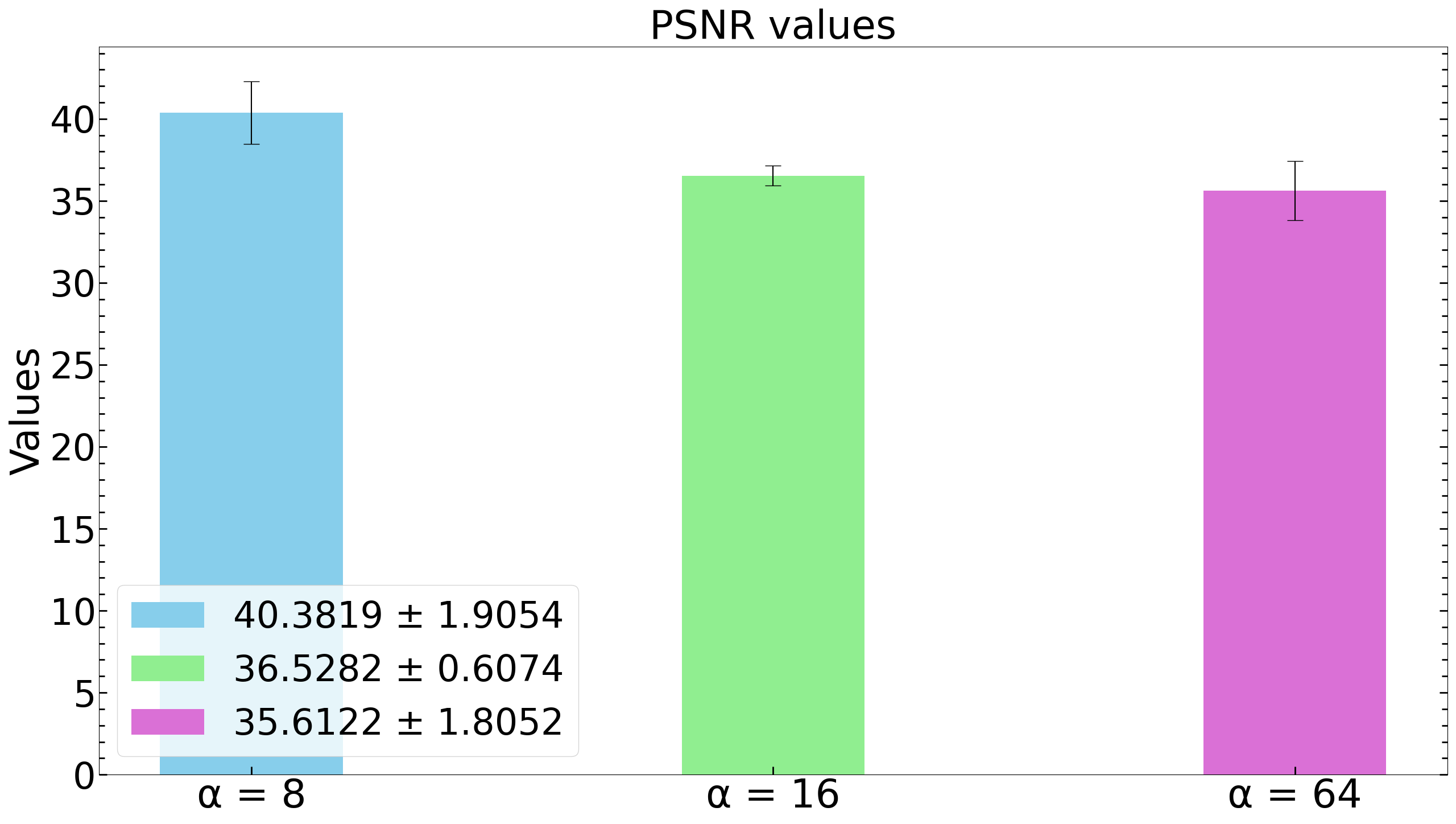}%
}\hspace{0.05\linewidth}
\subfloat[\label{fig:lpips}]{%
 \includegraphics[width=0.38\linewidth]{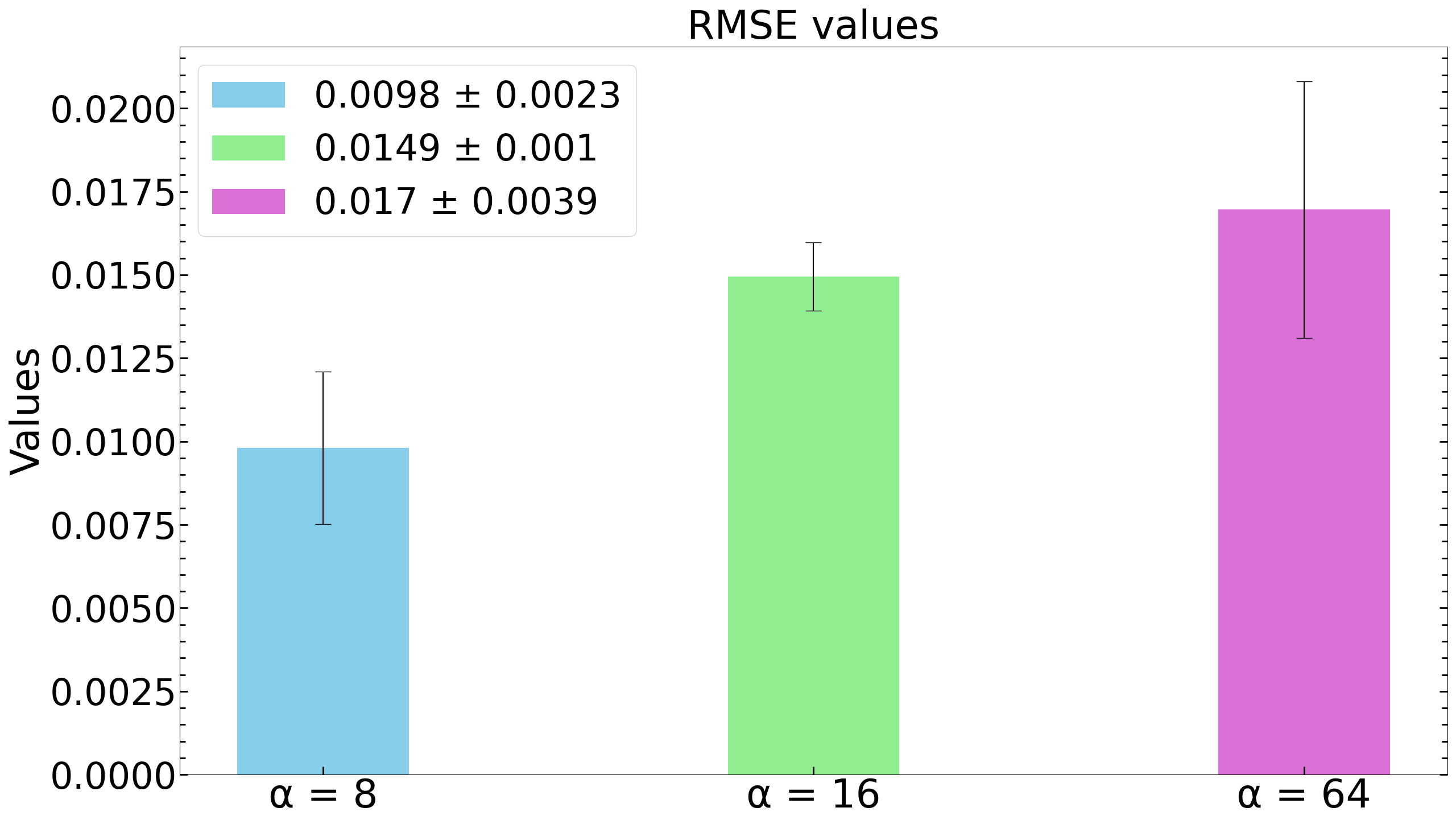}%
}\hspace{0.05\linewidth}
\subfloat[\label{fig:rmse}]{%
 \includegraphics[width=0.38\linewidth]{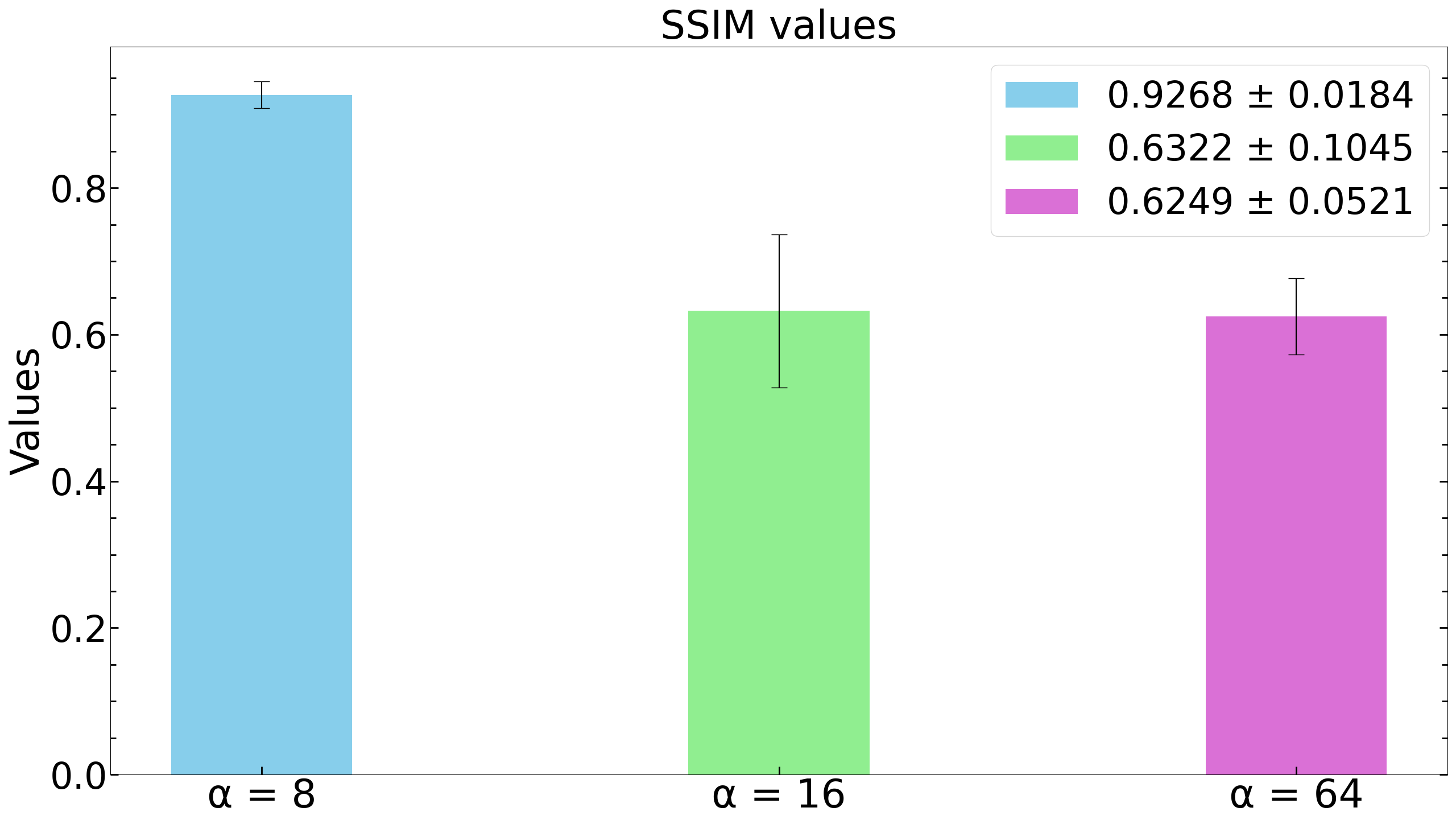}%
}

\caption{Comparison of image reconstruction metrics across varying downsampling factors \( \alpha = 8, 16, 64 \) using a VQGAN-based encoder-decoder architecture. The metrics include: (a) LPIPS to assess perceptual similarity, (b) PSNR for signal fidelity, (c) RMSE for pixel-wise error, and (d) SSIM for structural similarity. There is a noticeable trade-off between compression and image quality, with lower values of \( \alpha \) resulting in better perceptual and structural performance, while higher \( \alpha \) values lead to increased error and degradation in perceptual metrics.
}
\label{fig:ae_metrics}
\end{figure*}

The complexity of the SR tasks lies in their strongly ill-posed nature, as several HR images correspond to the same LR image. Traditional techniques use an average distance $\mathcal{L}$, such as the MSE in the pixel space, which leads to predicting an average of all the possible HR-predicted images struggling to replicate high-frequency details \citep{bruna2016superresolutiondeepconvolutionalsufficient, ledig2017photorealisticsingleimagesuperresolution}. DDPMs address this problem since the $\mathcal{L}$ is not related to the pixel domain \citep{ho2020, Ramunno2024}. Indeed, they do not predict the SR image directly, but the noise to be removed for obtaining the SR image. In addition, they are inherently probabilistic, enabling the possibility to model the probability distribution of all the possible HR images that correspond to the input LR image instead of predicting an average of those.
\subsection{Palette approach}
\label{sec:palette}

DDPMs \citep{ho2020} transform samples drawn from a standard Gaussian distribution into samples from the empirical data distribution through an iterative denoising process. Conditional diffusion models \citep{cfg} extend this approach by conditioning the denoising process on an input signal, thus enabling the generation of data based on the provided input. An example of conditional diffusion models are the Image-to-image diffusion models, which have been already applied for Super-Resolution tasks \citep{saharia2021imagesuperresolutioniterativerefinement, Saharia2021}.

During training, the actual HR image $y$ is used to generate a noisy version, denoted as $\tilde{y}$. This noisy version is then combined with the LR image $x$, and the resulting combined images are fed into the network to predict the noise added to the HR input image $y$. At inference time, the process starts with a pure Gaussian noise image, which is combined with the LR image $x$. The model is then used iteratively to produce the final predicted HR image. For further details, refer to the description in \cite{Saharia2021}.
\subsection{Latent Diffusion Model}
\label{sec:ldm}

\begin{figure*}
\centering
\subfloat[Ground Truth\label{fig:gt}]{%
 \includegraphics[width=0.45\linewidth]{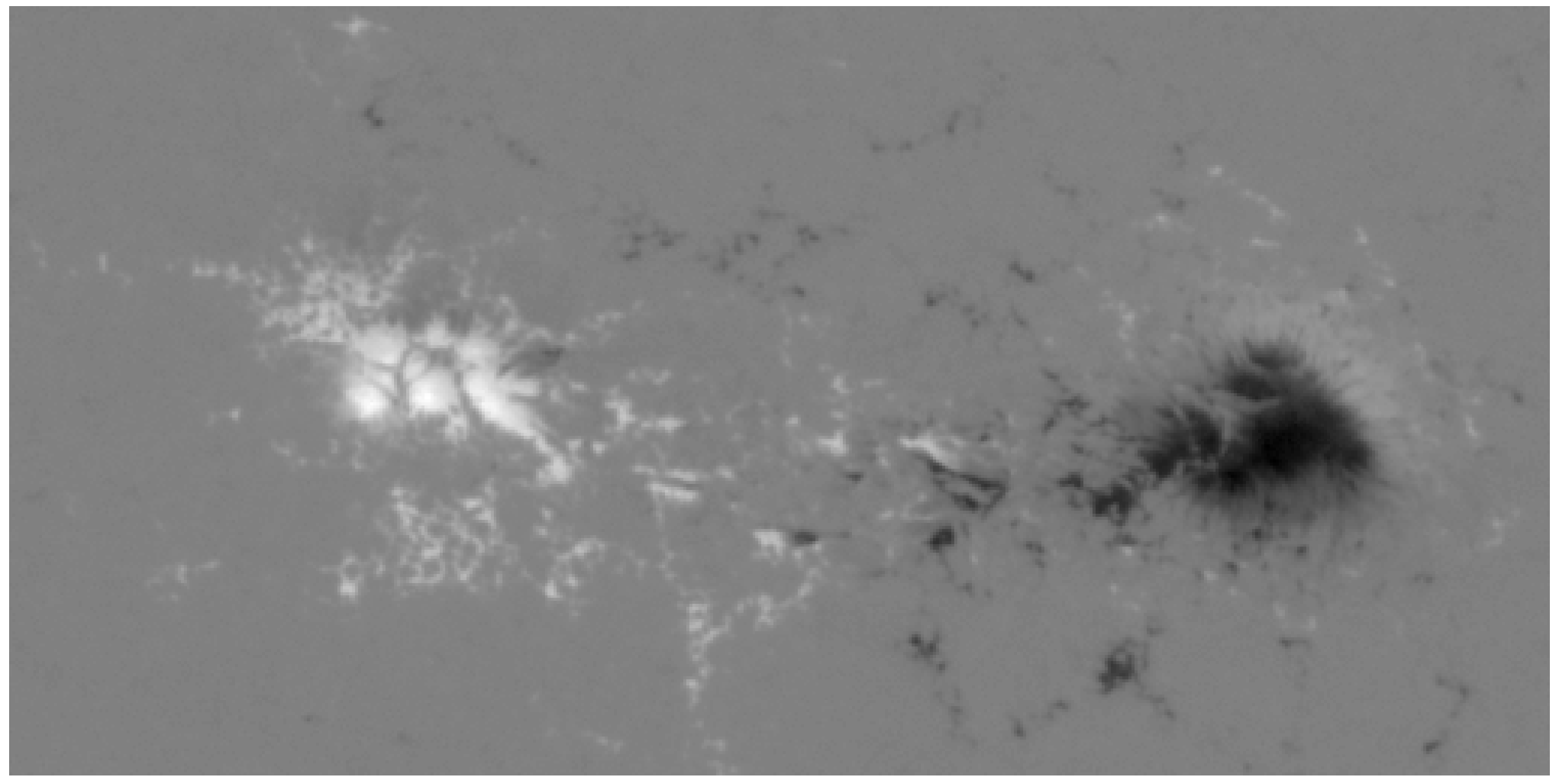}%
}\hspace{0.05\linewidth}
\subfloat[Input\label{fig:input}]{%
 \includegraphics[width=0.45\linewidth]{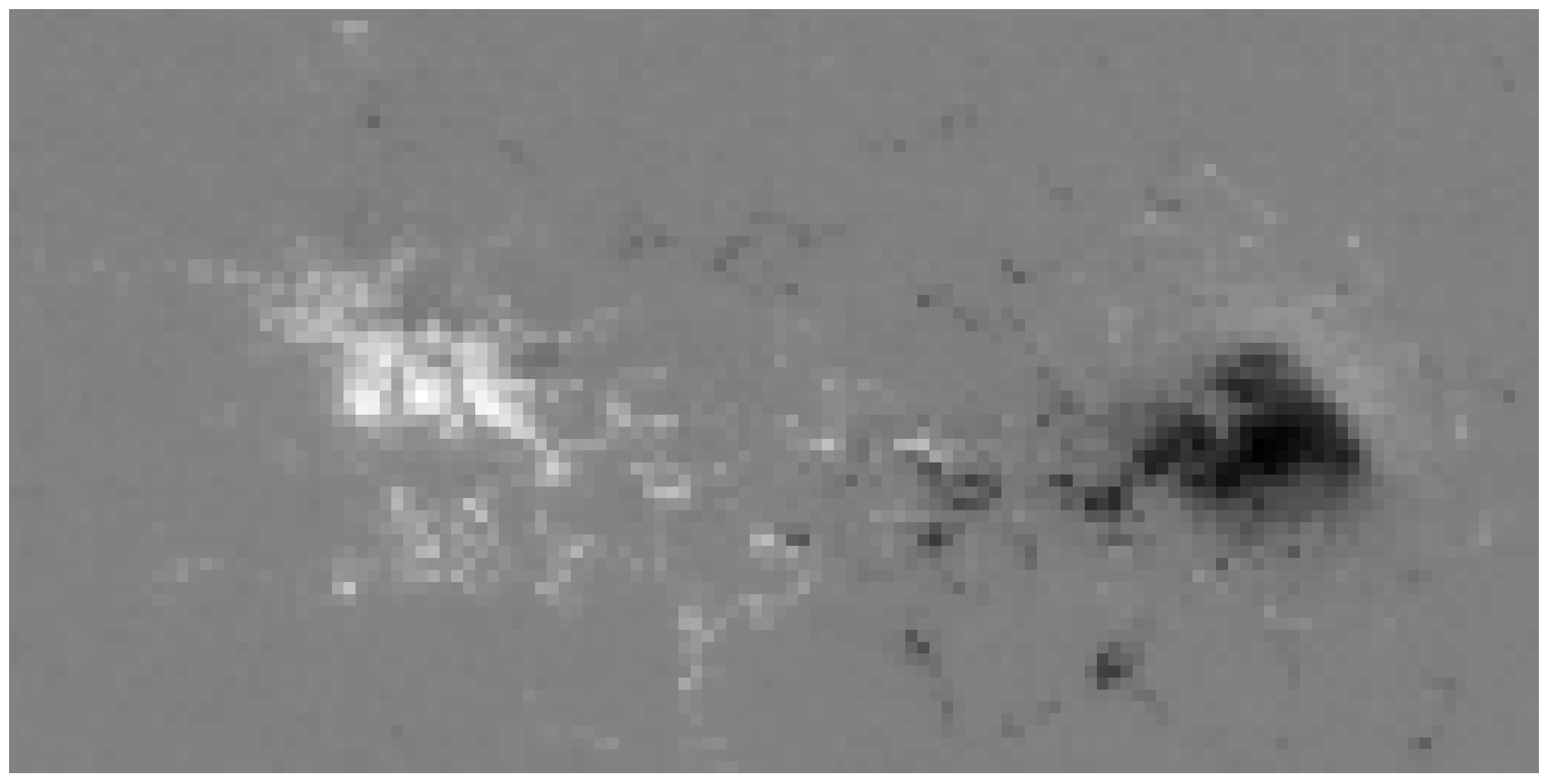}%
}\hspace{0.05\linewidth}
\subfloat[LDM with residuals\label{fig:ldmres}]{%
 \includegraphics[width=0.29\linewidth]{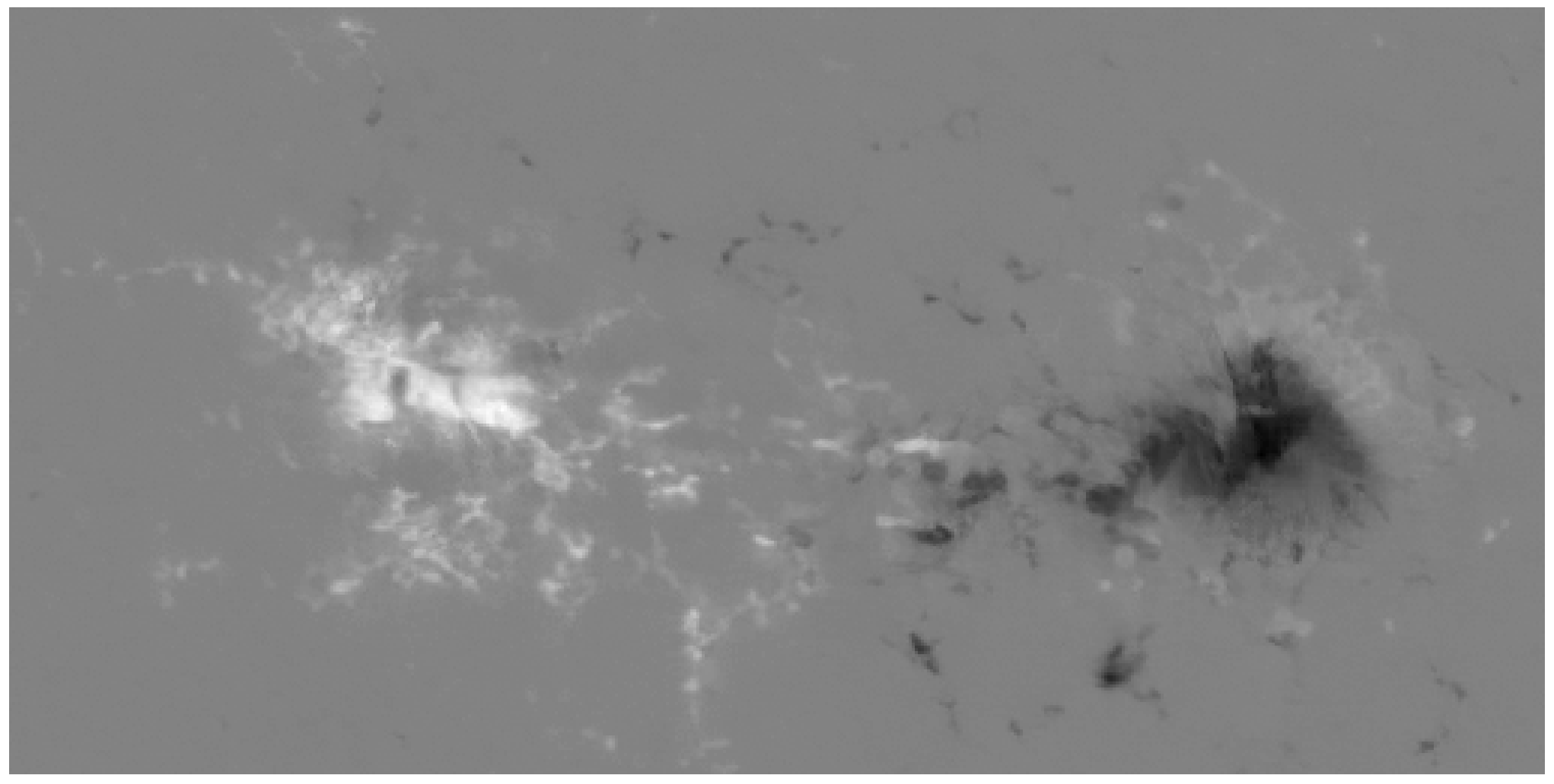}%
}\hspace{0.05\linewidth}
\subfloat[LDM\label{fig:ldm}]{%
 \includegraphics[width=0.29\linewidth]{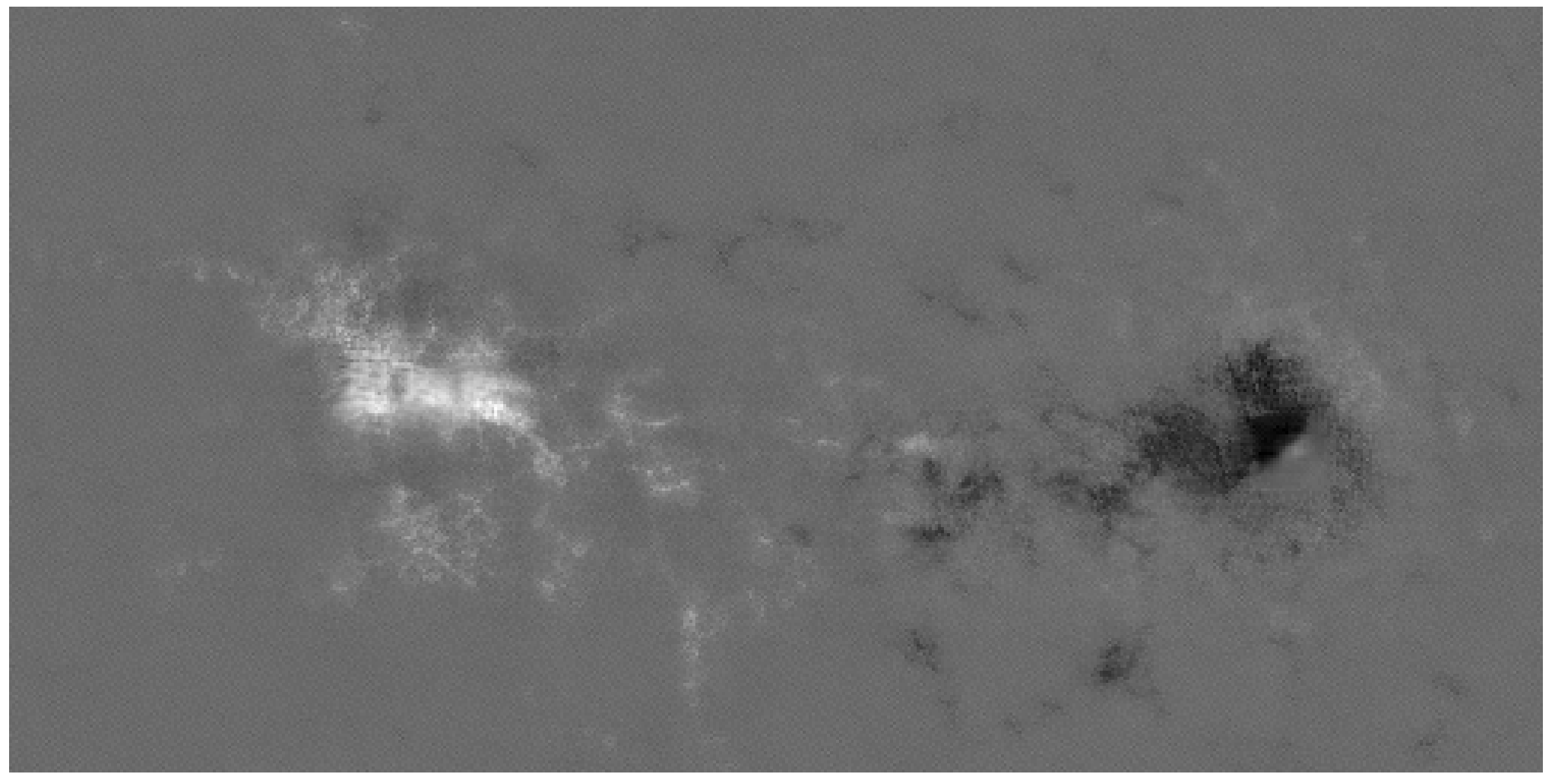}%
}\hspace{0.05\linewidth}
\subfloat[Palette with residuals\label{fig:paletteres}]{%
 \includegraphics[width=0.29\linewidth]{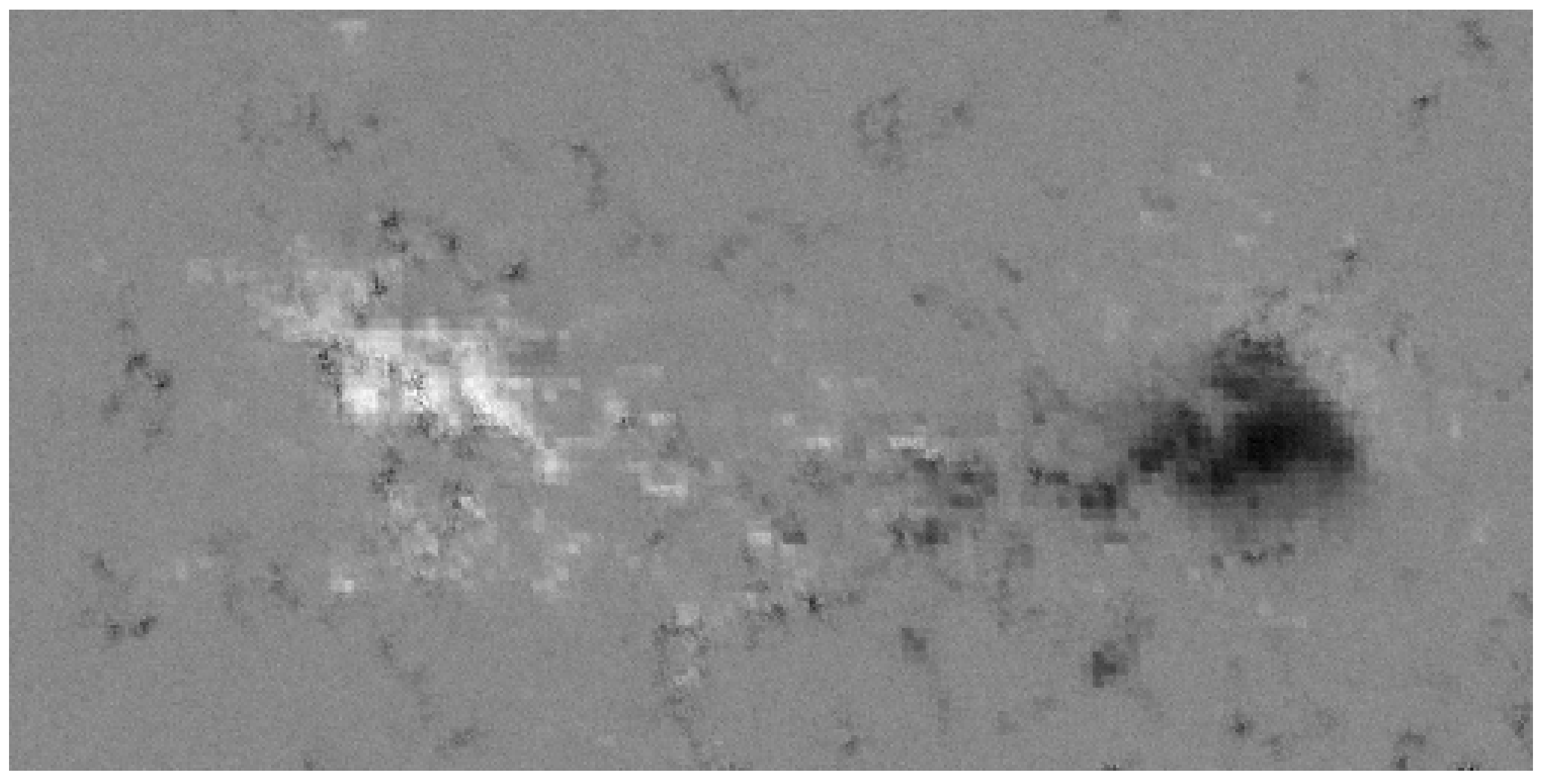}%
}\hspace{0.05\linewidth}
\subfloat[Palette \citep{Saharia2021}\label{fig:palette}]{%
 \includegraphics[width=0.29\linewidth]{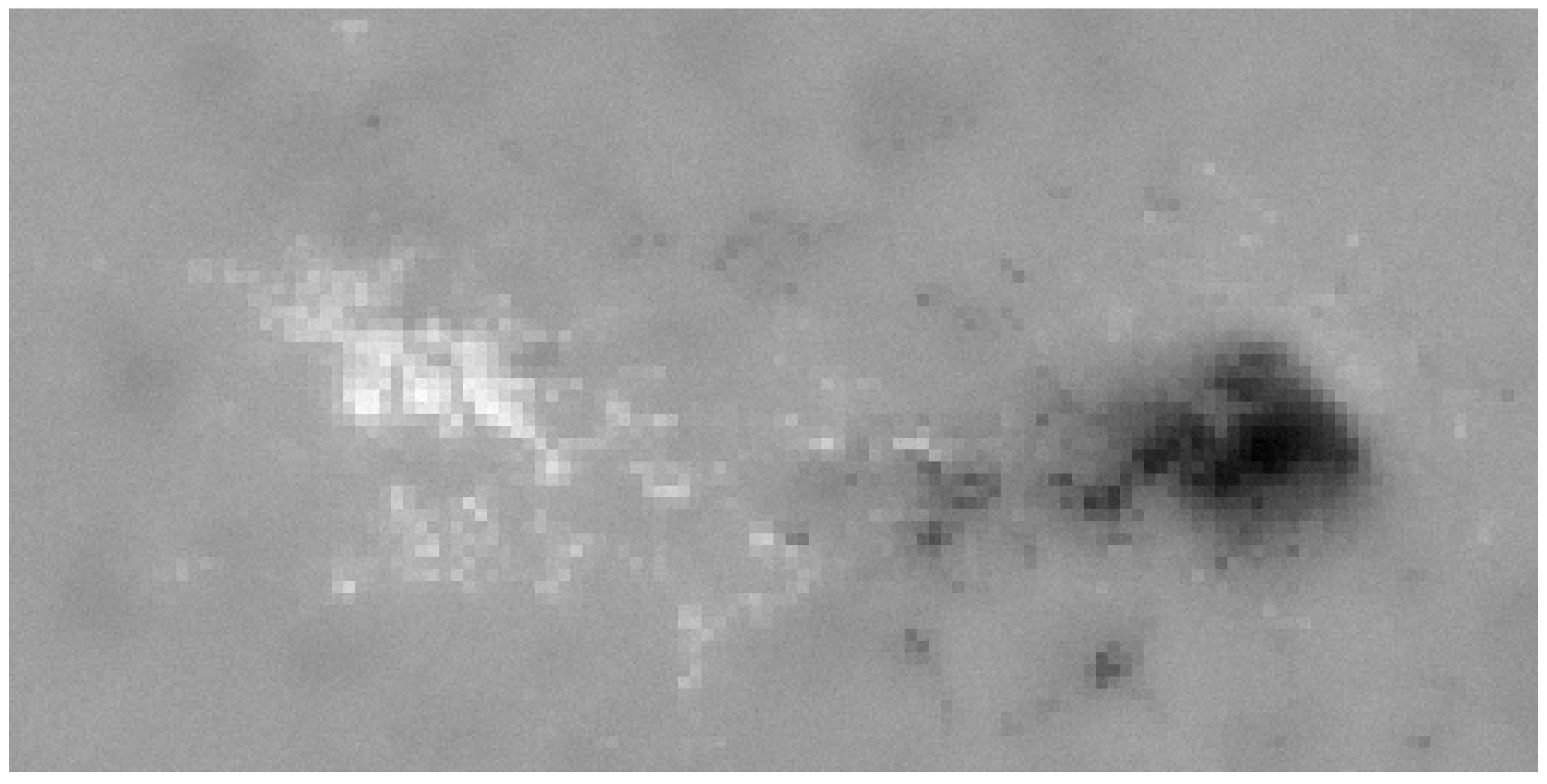}%
}\hspace{0.05\linewidth}
\subfloat[Enhance \citep{baso2018}\label{fig:enhance}]{%
 \includegraphics[width=0.29\linewidth]{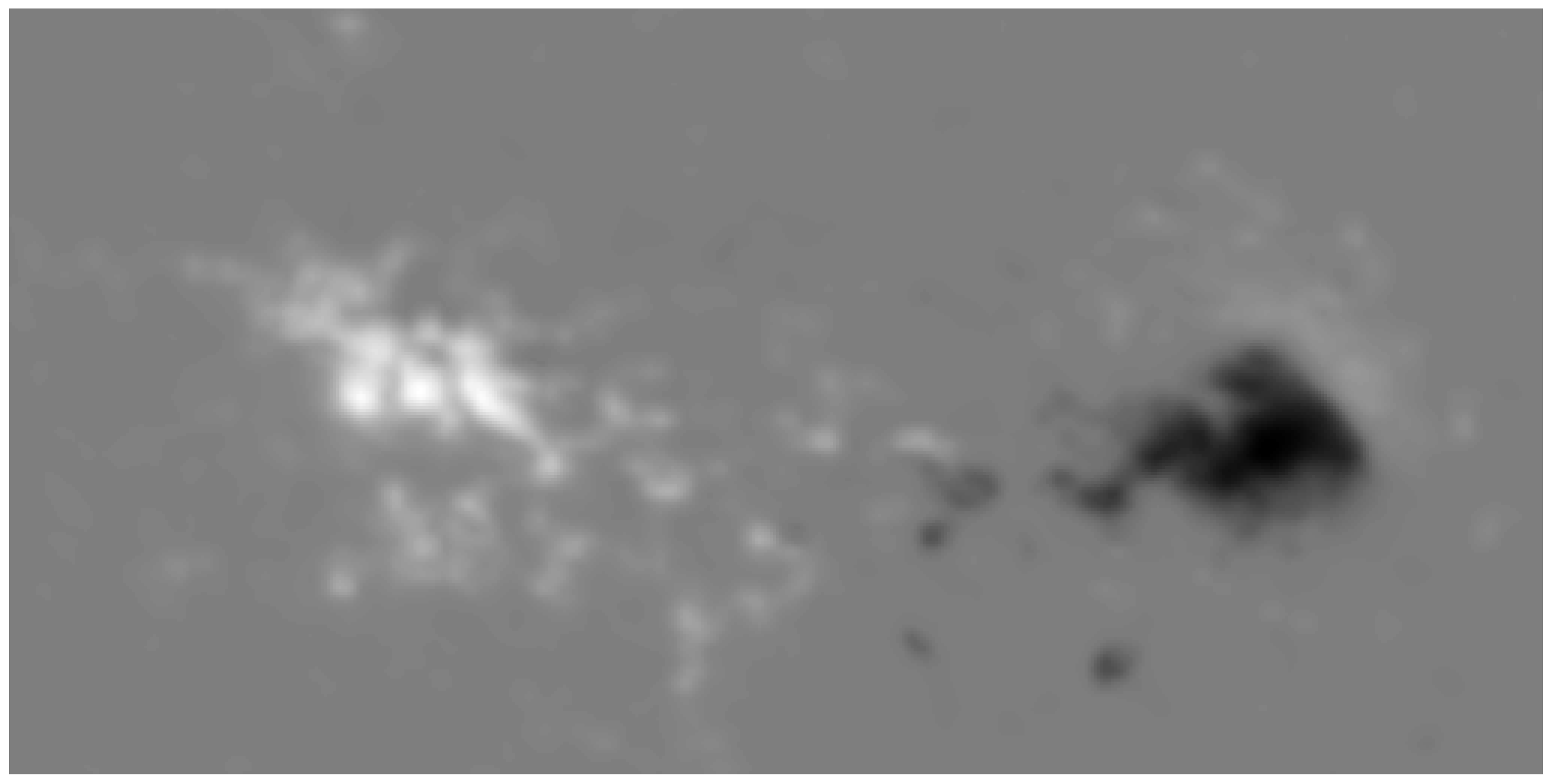}%
}\hspace{0.05\linewidth}
\subfloat[Progressive \citep{Rahman2020}\label{fig:progressive}]{%
 \includegraphics[width=0.29\linewidth]{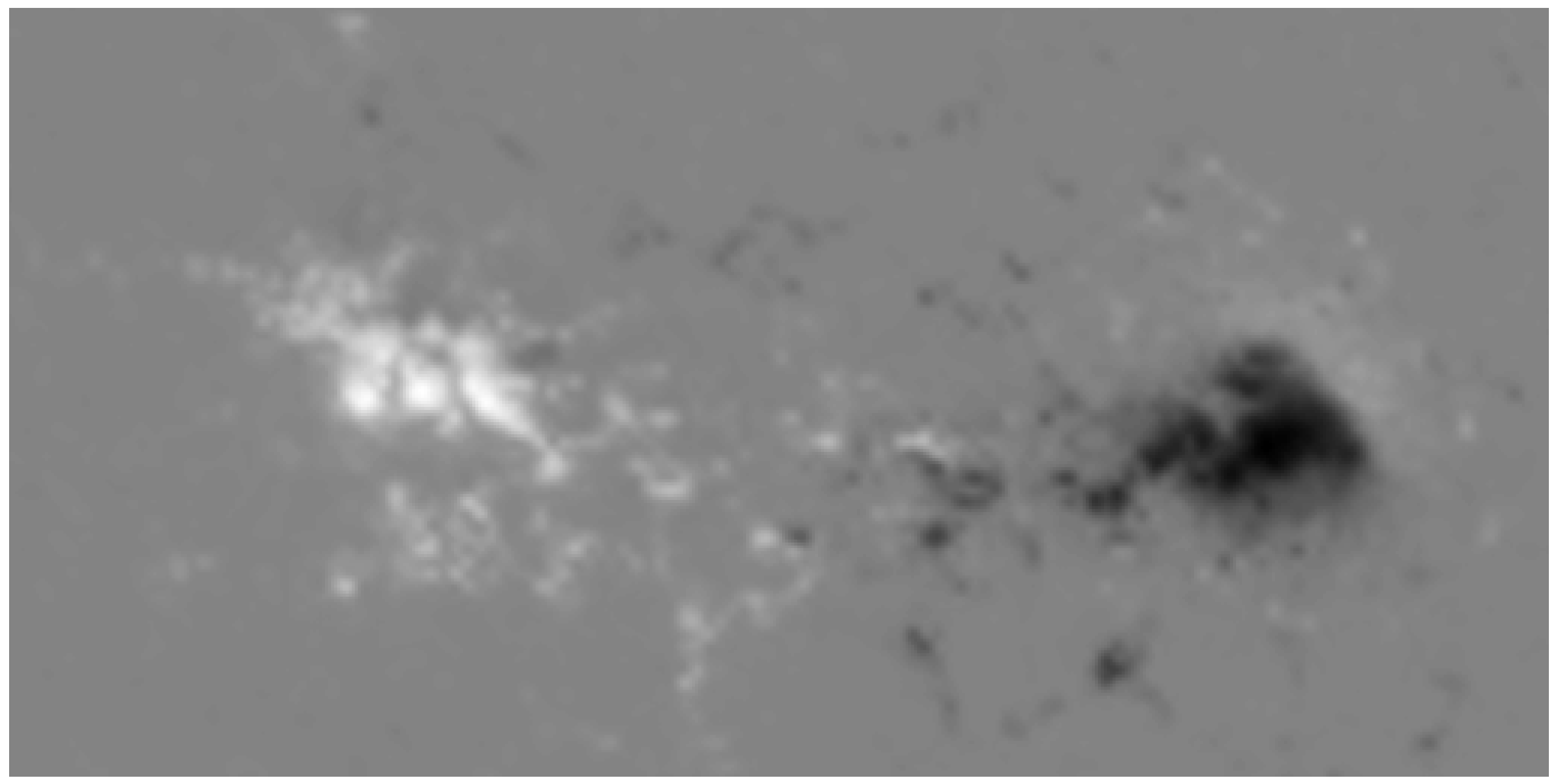}%
}
\caption{Comparison of different Super-Resolution techniques applied to the same HMI/SDO magnetogram. Panels (a) and (b) represent the ground truth and the degraded input image, respectively. The next images display the results of different methods: (c) Latent diffusion model with residuals, (d) Latent diffusion model without residuals, (e) Palette with residuals, (f) Palette without residuals, (g) Enhance model \citep{baso2018}, and (h) Progressive model \citep{Rahman2020}. This visual comparison highlights the impact of the LDM technique with residuals which shows fine-scale details with respect to the other reconstructions.
}
\label{fig:visual_inspections}
\end{figure*}

High resolution image synthesis is dominated by likelihood-based models \citep{ho2020, song2021scorebasedgenerativemodelingstochastic} and DDPMs have shown impressive results in the various domain of image synthesis, such as image generation, image-to-image translation and Super-Resolution \citep{saharia2021imagesuperresolutioniterativerefinement, kingma2023variationaldiffusionmodels}. These models, being based on likelihood estimation, avoid the issues of mode collapse and unstable training that are common in GANs \citep{Dhariwal2021}. However, the mode-covering nature of diffusion models, while effective in preventing mode collapse, often results in an over-allocation of capacity and computational resources to capturing imperceptible details. In addition, the need for up to thousands iterative steps during inference makes makes these models both slow and computationally expensive. 
To overcome this limitation, \cite{Rombach2021} presented for the first time the idea of training a DDPM not in the high-dimensional pixel space, 4096$\times$4096 pixels for the case of HMI data, but in a latent space of an already pre-trained autoencoder (AE). Since the AE model provides a lower-dimensional representational space which is perceptually equivalent to the data space, the diffusion model trained here are faster both in training and in inference with the possibility of having a more complex architecture backbone \citep{Rombach2021}. Additionally, the latent space distribution is potentially less complex and thus easier to model, even if equivalent to the data space distribution.
This variation of DDPMs is denoted as Latent Diffusion Model (LDM). In the Appendix \ref{sec:comp_time} there are more details about the computational time with respect the classica DDPM approach.

\subsection{Enhance and Progressive models}
In the past, various attempts have been made to achieve the Super-Resolution task from MDI to HMI. These approaches used classical neural networks with convolutional layers, which are deterministic by nature. Deterministic networks, in contrast to probabilistic models (e.g., diffusion models), produce a single, specific output for a given input without any randomness involved in the process.
We compare the probabilistic methods with two classical networks: Enhance by \cite{baso2018} and Progressive by \cite{Rahman2020}. We implemented and trained these networks from scratch. The input for both networks is a downsampled image of 256$\times$256 pixels, and the output is an image of 1024$\times$1024 pixels. For further architectural details, see Figure 3 in \cite{baso2018} for the Enhance model and Figure 1 in \cite{Rahman2020} for the Progressive model.

\section{Methodology and experiments}
\label{sec:experiments}
The experiments aim at finding the most suitable model to super-resolve LoS magnetograms from the MDI instrument to the spatial resolution of HMI. Specifically, we evaluate the reliability of features smaller than 2'', which cannot be imaged by MDI.

The backbone of our diffusion model architectures \citep{ho2020, Rombach2021} consists of a U-Net \citep{ronneberger2015unetconvolutionalnetworksbiomedical}, which is an encoder-decoder network with skip connections where the input shape and the output shape are the same.
We train the models for a total of 30 epochs each using the AdamW \citep{loshchilov2019decoupledweightdecayregularization} optimizer, the MSE as loss function, a learning rate of $3 \times 10^{-4}$, a batch size of 4 and one NVIDIA TITAN X graphics processing unit (GPU). The model is implemented with the PyTorch framework \citep{pytorch}. We use mixed precision during training.

Regarding hyperparameter selection, we did not employ automated optimization tools such as grid search or random search due to the high computational cost and the empirical nature of tuning diffusion models. Instead, we selected key hyperparameters like the learning rate, batch size, and number of epochs based on prior research and manual tuning. Specifically, we followed the hyperparameter choices outlined by \cite{ho2020}, which are widely adopted for denoising diffusion probabilistic models. These settings have been shown to yield stable and reliable results for this class of models.

Additionally, we experimented with dynamic learning rate strategies, such as cosine annealing and learning rate warm-up. However, these approaches did not improve performance compared to our fixed learning rate of $3 \times 10^{-4}$, which resulted in more stable convergence and overall better performance.
To monitor and track the training process, we utilized Weights and Biases \citep[Wandb]{wandb}, enabling us to visualize learning curves in real time and apply early stopping to avoid overfitting.

The input and ouput images of our model have a size of 1024$\times$1024 pixels. To create input images with a spatial resolution of 2''/pixel, we downgrade the HMI images as follows. We randomly crop a 1024$\times$1024-pixel region from a 4096$\times$4096-pixel full-disk LoS magnetogram. Then, we compute the average value over every 4$\times$4-pixel block, resulting in an image of 256$\times$256 pixels. Finally, we upscale this image to a size of 1024$\times$1024 pixels by replicating each pixel 4 times without interpolation. This process simulates a 2"/pixel resolution crop of the HMI data, with the same image size as the original 1024x1024 crop at 0.5"/pixel resolution (Figure \ref{fig:downscaled}).

We train 4 diffusion model frameworks and 2 deterministic models with the same backbone architectures as presented in \cite{baso2018} (Enhance) and \cite{Rahman2020} (Progressive). All the models are trained with the same input data, the difference among the 4 diffusion frameworks is based on the usage of DDPM in the pixel space \citep{ho2020, Ramunno2024, Ramunno2024mag2mag} or in the latent space \citep{Rombach2021} and on the choice of predicting the difference of the HR and the LR image or directly the HR image.

We aim to evaluate whether adopting a probabilistic approach provides more advantages compared to a deterministic one by comparing diffusion model frameworks with deterministic frameworks. Furthermore, we want to explore if working in the latent space offers significant benefits over operating directly in the pixel space. To do this, we analyze two frameworks that function in pixel space using the Palette technique \citep{Saharia2021}, as described in Section \ref{sec:palette}, and two other frameworks where the data is first encoded using a pre-trained Autoencoder (AE) before applying the Palette technique in the latent space.
The pre-trained autoencoder is provided by the Hugging Face Diffusers library \citep{diffusers}. This model is a Vector Quantized Generative Adversarial Network (VQGAN) \citep{esser2021tamingtransformershighresolutionimage} and we use the pre-trained weights from Hugging Face and the work by \cite{Rombach2021}. Additionally, we train the LDM on the non-quantized latent space.
For both latent space and pixel space generation approaches, we investigate whether it is more effective to predict the HR image directly or to predict the residual information between the LR and HR images (i.e., the difference). As demonstrated in prior work \citep{li2021srdiffsingleimagesuperresolution, whang2021deblurringstochasticrefinement}, focusing on residual details allows DMs to concentrate on finer features, which accelerates convergence and stabilises training. A sketch of the training step and the inference algorithm of our network is presented in Figure \ref{fig:concept_and_architecture} and the architecture backbone is given in Figure \ref{fig:architecture}.
In Figure \ref{fig:training} we encode the downscaled image and the target image, then we compute their difference (residual) in the latent space and perturb the residual with gaussian noise with a magnitude determined by the timestep t, a discrete parameter that varies between 0 and 1000. Afterward, we concatenate channel-wise the noised residual with the encoded downscaled image and we pass it through the U-Net backbone in Figure \ref{fig:architecture} with the aim of predicting the injected noise with the MSE loss.
\begin{table*}
    \centering
    \begin{tabularx}{\textwidth}{|l|X|X|X|X|X|X|}
    \hline
        \textbf{Metric} & \textbf{LDM RES (Ours)} & \textbf{LDM NO RES \citep{Rombach2021}} & \textbf{DDPM RES (Ours)} & \textbf{DDPM NO RES \citep{Saharia2021}} & \textbf{Enhance \citep{baso2018}} & \textbf{Progressive \citep{Rahman2020}} \\ \hline
        PSNR $\uparrow$ & 38.2 ± 2 & 29.2 ± 5 & 21.8 ± 6.6 & 17.2 ± 7.5 & 37.6 ± 4 & \textbf{38.8 ± 4} \\ \hline
        SSIM & 0.9 ± 0.03 & 0.23 ± 0.21 & 0.14 ± 0.2 & 0.06 ± 0.13 & 0.92 ± 0.01 & \textbf{0.94 ± 0.01} \\ \hline
        LPIPS $\downarrow$ & \textbf{0.03 ± 0.01} & 0.09 ± 0.04 & 0.63 ± 0.21 & 0.47 ± 0.17 & 0.19 ± 0.04 & 0.15 ± 0.02 \\ \hline
        FID $\downarrow$ & \textbf{0.01 ± 0.0} & 0.38 ± 0.02 & 1.76 ± 0.12 & 3.0 ± 0.56 & 0.02 ± 0.01 & 0.02 ± 0.01 \\ \hline
        Unsigned Magnetic Flux (\%) & \textbf{7.52 ± 8.56} & 58.3 ± 27.9 & 73.3 ± 26.4 & 81.0 ± 25.0 & 51.1 ± 34.6 & 17.7 ± 12.0 \\ \hline
        AR Size (\%) & \textbf{8.0 ± 11.1} & 17.7 ± 19.3 & 67.8 ± 36.3 & 80.4 ± 36.6 & 28.5 ± 58.5 & 9.0 ± 16.7 \\ \hline
    \end{tabularx}
    \caption{Comparison of metrics for different super-resolution methods.}
    \tablefoot{
        The symbol $\downarrow$ indicates that a lower value is preferable for the metric, while the symbol $\uparrow$ indicates that a higher value is preferable. The SSIM has values between -1 and 1. The unsigned magnetic flux and AR size are expressed as percentage variations. References: \citet{Rombach2021}, \citet{Saharia2021}, \citet{baso2018}, \citet{Rahman2020}.
    }
    \label{tab:results_metrics}
\end{table*}
\section{Metrics}
\label{sec:metrics}

To comprehensively assess our model's performance, we define two distinct sets of evaluation metrics. The first set originates from the computer science domain and focuses on measuring the visual quality of the generated images. The second set comes from the physics domain, evaluating the physical accuracy and reliability of the super-resolved images to ensure that that they adhere to real-world physical principles.
The first set comprises the Peak-Signal-to-Noise-Ratio (PSNR), the Structural-Similarity-Index Measure (SSIM) \citep{ssim}, the Learned Perceptual Image Patch Similarity (LPIPS) \citep{lpips} and the Fréchet inception distance (FID) \cite{fid}.
The PSNR measures the ratio between the maximum possible power of a signal (the image) and the power of the noise that distorts the signal and is defined as:
\begin{equation}
    \text{PSNR} = 10 \cdot \log_{10} \left( \frac{MAX_{y}^2}{\text{MSE}} \right),
\end{equation}
where
\begin{equation}
    \text{MSE} = \frac{1}{w \cdot h} \sum_{i=0}^{w-1} \sum_{j=0}^{h-1} \left( y(i,j) - \hat{y}(i,j) \right)^2,
\end{equation}
where \( y \) represents the ground truth HR image and \( \hat{y} \) represents the predicted HR image. The higher the PSNR, the better the reconstruction quality.
The SSIM measures the similarity between two images based on three components: luminance, contrast, and structure.
\begin{equation}
    \text{SSIM}(y, \hat{y}) = \frac{(2\mu_y \mu_{\hat{y}} + C_1)(2\sigma_{y\hat{y}} + C_2)}{(\mu_y^2 + \mu_{\hat{y}}^2 + C_1)(\sigma_y^2 + \sigma_{\hat{y}}^2 + C_2)},
\end{equation}
where \( \mu_y \) and \( \mu_{\hat{y}} \) are the mean intensities, \( \sigma_y^2 \) and \( \sigma_{\hat{y}}^2 \) are the variances, and \( \sigma_{y\hat{y}} \) is the covariance between \( y \) and \( \hat{y} \).
The constants $C_1$ and $C_2$ are used to stabilize the formula. SSIM values range from -1 to 1, where 1 indicates that the images are identical in terms of structural similarity.

It is known that PSNR and SSIM tend to favor blurry images \citep{Dahl2017, Ledig2017, menon2020}, meaning that models producing such outputs can achieve high scores in these metrics. This is misleading as it does not accurately reflect the true quality of model performance. The underlying cause of this issue is typically linked to the loss function used during training, which is often MSE. MSE encourages the model to predict the average of possible outcomes rather than the most precise or sharpest result, leading to smoother, less detailed predictions. 

In our case, although MSE appears in the training process, its role is fundamentally different compared to classical super-resolution models. Specifically, in diffusion-based super-resolution models, MSE is used at each denoising timestep to predict the added noise, rather than directly optimizing the final super-resolution output in pixel space. This distinction allows the model to probabilistically model the distribution of possible high-resolution outputs, preserving high-frequency details and reducing the risk of overly smooth predictions. Furthermore, while classical models produce the output in a single step, diffusion models iteratively refine the prediction over multiple steps, progressively improving detail quality and mitigating the averaging problem. Moreover, MSE loss has been shown to be optimal for DDPM training for added noise prediction \citep{ho2020} and also the encoder-decoder architecture used is suited for MSE loss.

Therefore, to better evaluate the model's performance, we use the LPIPS distance, an L2 norm in the latent space of a pre-trained AlexNet model \citep{alexnet}. LPIPS is valuable because it focuses on perceptual features that humans consider when evaluating image quality, making it more effective than traditional metrics to assess image quality in generative models. 

To support the results obtained with the LPIPS distance, we use the Fréchet Inception Distance (FID). FID measures how similar the distribution of generated images is to the distribution of real images by comparing their statistical properties. It is calculated as the distance between the multivariate Gaussian distributions of real and generated images in the latent space of a pre-trained encoder CLIP \citep{clip}:

\begin{equation}
    \text{FID} = \left\| \mu - \hat{\mu} \right\|^2 + \text{Tr}\left( \Sigma + \hat{\Sigma} - 2 \sqrt{\Sigma \hat{\Sigma}} \right),
\end{equation}

where \( \mu \) and \( \hat{\mu} \) are the means of real and generated images. \( \Sigma \) and \( \hat{\Sigma} \) are the covariance matrices of the real and generated images, and \( \text{Tr} \) represents the trace of the matrix.

Nevertheless, our primary focus is on the physical accuracy of the super-resolved magnetograms rather than solely their human-perceived aesthetics. For this reason, we use metrics such as unsigned magnetic flux and the size of the active regions. These metrics are evaluated in terms of percentage variations relative to the corresponding values in the ground-truth magnetograms. Ideally, the closer the percentage variations to 0\%, the better the performance of the model. Before calculating these metrics, the images are first brought back to their original range in Gauss. We then use the SunPy library \citep{sunpy_community2020} to identify the centres of the active regions (ARs) through their NOAA (National Oceanic and Atmospheric Administration) numbers for accurate localisation.
To compute the unsigned and the net flux we sum over the pixel in Gauss and then multiply by the total area considered. For the active region size we use the OpenCV library \citep{opencv_library} to compute the contour of the ARs. To do this, we first binarize the image setting as a threshold 300 G, then we find the contours via the findContours() function from the OpenCV library and then we count the number of pixels inside to calculate the approximation of the area in units of number of pixels.

\section{Results and Discussion}
\label{sec:results}

Our first experiment focuses on identifying the most effective model for performing Super-Resolution on LoS magnetograms. To achieve this, it is essential to isolate the Super-Resolution task in a controlled environment where the only distinction between the LR and HR images is their spatial resolution. Therefore, we conduct this experiment using data obtained from the HMI/SDO instrument.

\begin{figure*}
\centering
\subfloat[MDI (Input)\label{fig:mdi}]{%
 \includegraphics[width=0.40\linewidth]{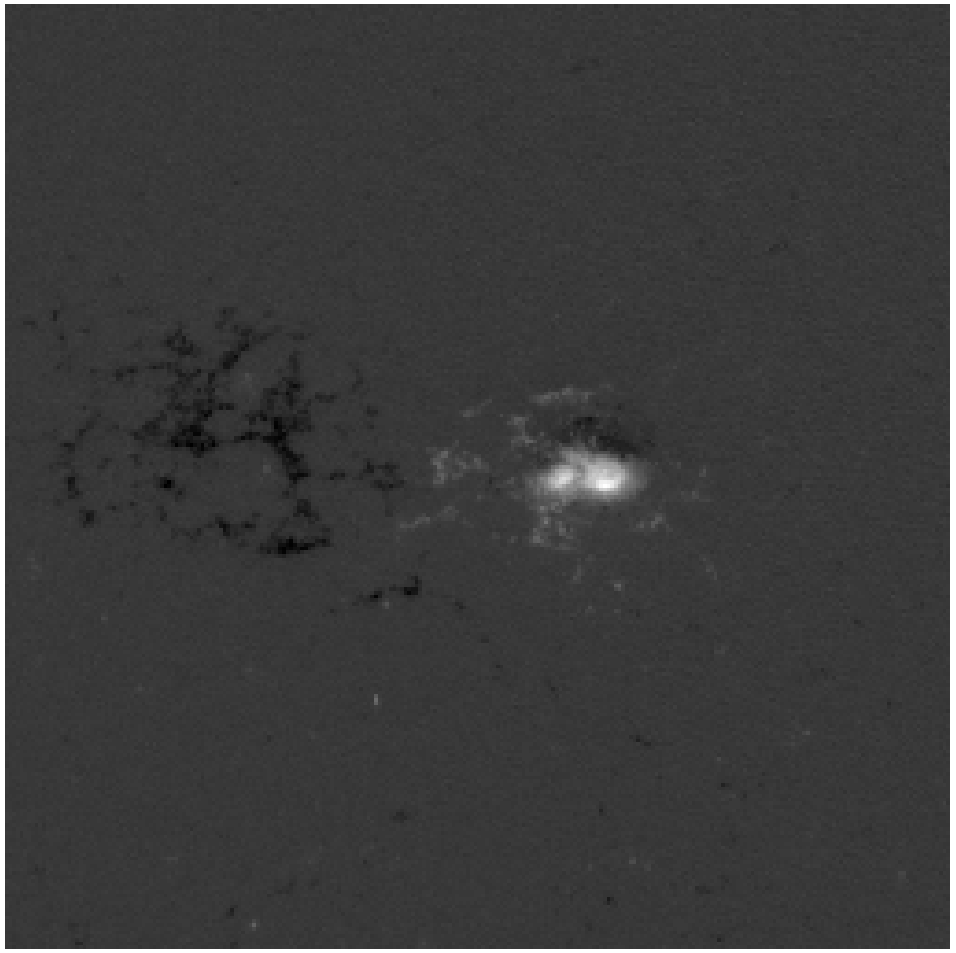}%
}\hspace{0.05\linewidth}
\subfloat[HMI (Ground Truth)\label{fig:hmi}]{%
 \includegraphics[width=0.40\linewidth]{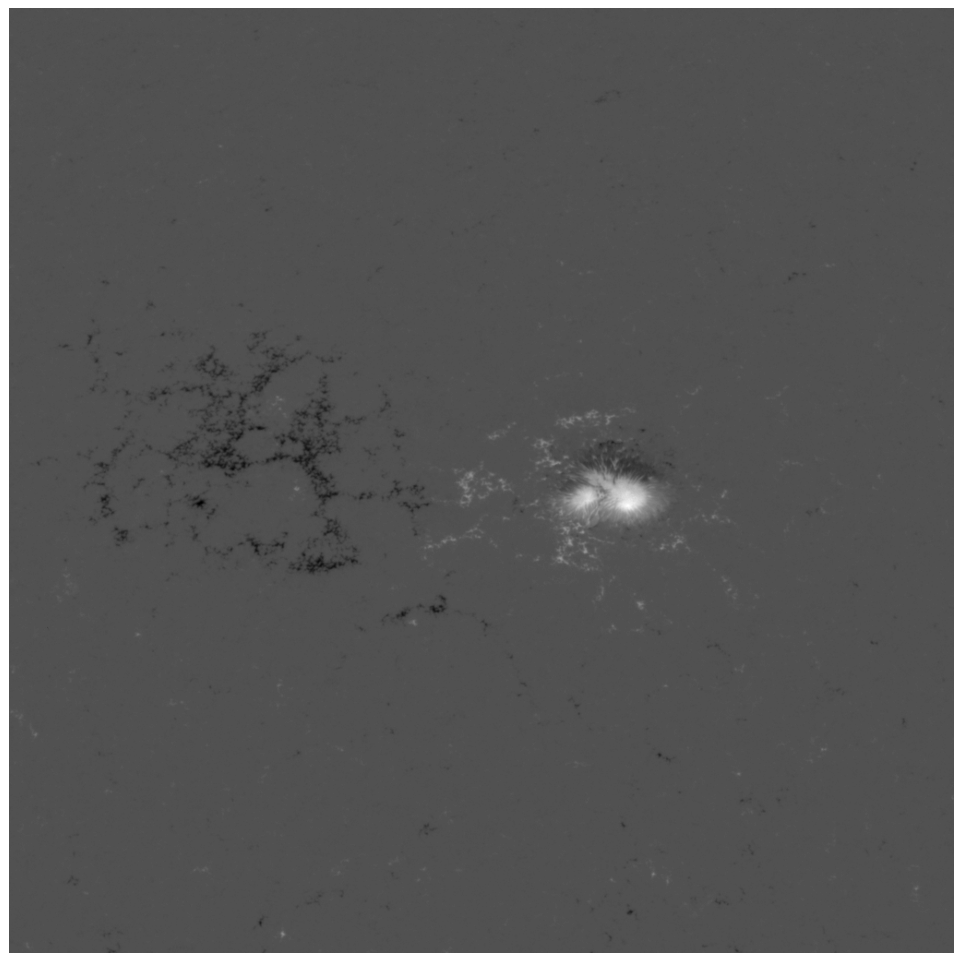}%
}
\caption{Comparison of an MDI and HMI observation. NOOA AR number 11108 from 22 September 2010.}
\label{fig:mdi_hmi}
\end{figure*}

Since our goal is to super-resolve MDI/SOHO magnetograms to match the resolution of HMI/SDO magnetograms, we must train a model capable of performing 4x Super-Resolution. This is because we need to enhance the spatial resolution from 2"/pixel to 0.5"/pixel. We train all the models on aligned crops of LR and HR magnetograms, thus for each HMI image we take a random crop of 1024 $\times$ 1024 pixels, we downsample it to 256 $\times$ 256 pixels by averaging every 4 pixels and then for the diffusion model frameworks we upsample it again by just replicating each pixel 4 times leading to a final image size of 1024 $\times$ 1024 pixels (Figure \ref{fig:downscaled}), while for the Enhance and Progressive approach the input image has a image size of 256 $\times$ 256 pixels.
For the diffusion framework we upsample it again before passing through the model because we are using the U-Net \citep{ronneberger2015unetconvolutionalnetworksbiomedical} and the input size and the output size must match. 

The architecture of the proposed network is illustrated in Figure \ref{fig:architecture}. This network integrates the latent diffusion approach \citep{Rombach2021} with the residual learning strategy \citep{li2021srdiffsingleimagesuperresolution}. By combining these approaches, we gain the computational efficiency of the latent space representation while also benefiting from the residual domain, where the image is mostly zero except in areas with important details. This effectively guides the model to focus on the regions that matter most for accurate prediction.

To train a DDPM in the latent space, it is necessary to have an already trained encoder-decoder architecture. We use the pre-trained networks from \cite{Rombach2021}. Specifically, the network used is the VQGAN \citep{esser2021tamingtransformershighresolutionimage}, which is the VQVAE (Vector-Quantized Variational Autoencoder) with the addition of a patch-wise discriminator loss.
The VQGAN employs three losses to stabilize the latent space and avoid arbitrarily high variance. The first is the use of the reconstruction loss, ensuring that the decoded images from the VQGAN generator are close to the original images. The second involves discretizing the latent space representation with the codebook loss, that aligns the encoder outputs with specific points in a predefined set of discrete vectors, called the codebook. Finally, the patch-wise discriminator loss is used to enhance image quality and make reconstructions more realistic. This improves both image generation quality and compression efficiency \citep{Oord2017, esser2021tamingtransformershighresolutionimage, Rombach2021}. For more details about the VQGAN training process we refer the reader to the work by \cite{esser2021tamingtransformershighresolutionimage}.

Given an image $x \in \mathbb{R}^{\hat{h} \times \hat{w} \times c}$, the encoder $\varepsilon$ encodes $x$ into a latent representation $z = \varepsilon(x)$, where $z \in \mathbb{R}^{h_{z} \times w_{z} \times c_{z}}$. The encoder downsamples the image by a factor $\alpha$, such that $\alpha = \hat{h} / h_{z} = \hat{w} / w_{z}$. We test three different downsampling factors $\alpha = 8, 16, 64$.

We test the autoencoder networks on the training set described in Section \ref{sec:data_source}. As shown in Figure \ref{fig:ae_metrics}, there is a clear trend where lower values of $\alpha$ result in better performance, while higher values of $\alpha$ lead to increased errors, consistent with \citep{Rombach2021}. Given that we are working with input images of spatial dimensions 1024 $\times$ 1024 pixels and prioritise quality, we choose the best-performing model with $\alpha = 8$.

To demonstrate that our network benefits from both the residual and latent diffusion techniques, we compare its performance by training with and without the latent space approach, as well as with and without the residual approach, as outlined in Section \ref{sec:experiments}.

\begin{table*}
    \centering
    \begin{tabularx}{\textwidth}{|l|X|X|X|}
    \hline
        \textbf{Metric} & \textbf{LDM RES (Ours)} & \textbf{Enhance \citep{baso2018}} & \textbf{Progressive \citep{Rahman2020}} \\ \hline
        PSNR $\uparrow$ & 29.23 ± 8 & \textbf{30.02 ± 9} & 29.71 ± 8  \\ \hline
        SSIM & 0.8 ± 0.05 & \textbf{0.9 ± 0.03} & 0.8 ± 0.04  \\ \hline
        LPIPS $\downarrow$ & \textbf{0.08 ± 0.03} & 0.16 ± 0.04 & 0.16 ± 0.04 \\ \hline
        FID $\downarrow$ & \textbf{0.04 ± 0.01} & 0.7 ± 0.2 & 2.3 ± 0.3  \\ \hline
        Unsigned Magnetic Flux (\%) & \textbf{18 ± 13} & 27.5 ± 26.7 & 28.5 ± 15.0 \\ \hline
        AR Size (\%) & \textbf{25 ± 18} & 35 ± 69 & 36 ± 38 \\ \hline
    \end{tabularx}
    \caption{Comparison of MDI to HMI super-resolution results.}
    \tablefoot{
        The symbol $\downarrow$ indicates that a lower value is preferable for the metric, while $\uparrow$ indicates that a higher value is preferable. SSIM ranges between -1 and 1. Unsigned magnetic flux and AR size are expressed as percentage variations. References: \citet{baso2018}, \citet{Rahman2020}.
    }
    \label{tab:results_mdi2hmi}
\end{table*}

The complexity of the architecture differs among the four diffusion frameworks, depending on whether we are training a latent diffusion model or working in the pixel domain. In our network, a key hyperparameter controls the number of channels in the convolutional layers, directly affecting model complexity. A larger value increases the number of parameters, enhancing the model capacity to learn complex features, but it also results in higher memory consumption and longer computation times. Conversely, a smaller value reduces the memory and computational load, making the model more lightweight, but potentially limiting its ability to capture intricate details, which may reduce performance in tasks requiring high accuracy.
For the latent diffusion approach, we use a downsampling factor of $\alpha = 8$, which reduces the spatial dimensions to 128 $\times$ 128 pixels after compression. In this case, we can set the aforementioned hyperparameter to 128, while in the pixel domain with images of 1024 $\times$ 1024 pixels, we are constrained to using a maximum of 8 due to computational limitations.
Moreover, in the latent diffusion approach, we can alternate between convolutional and self-attention layers \citep{vaswani2023attentionneed}. Self-attention allows the model to capture global relationships between pixels, model long-range dependencies, and aggregate features across the image. In Super-Resolution tasks, this helps the network concentrate on challenging regions and synthesise details accurately \citep{SU2025110935}. However, due to computational constraints, self-attention layers cannot be used in the pixel domain.

\begin{figure*}
\centering
\subfloat[Target HMI\label{fig:gt_zoom_crop}]{%
 \includegraphics[width=0.58\linewidth]{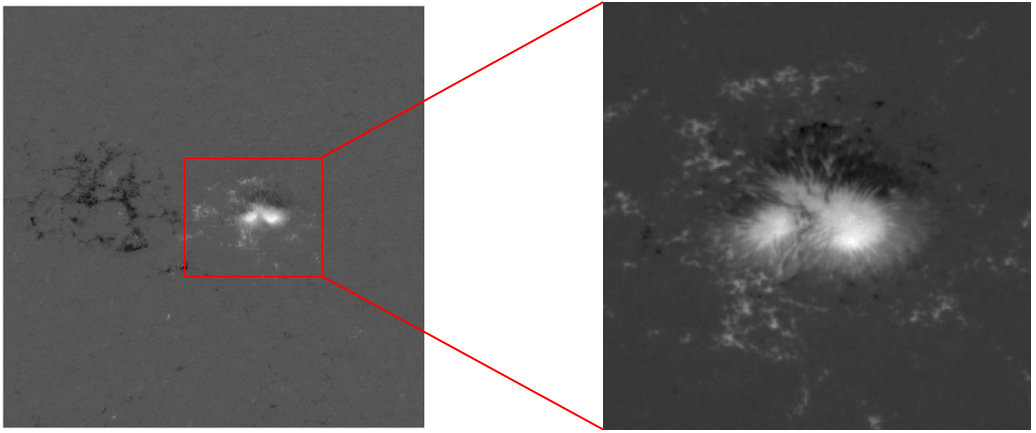}%
}\hspace{0.05\linewidth}
\subfloat[Input MDI\label{fig:mdi_zoom_crop}]{%
 \includegraphics[width=0.58\linewidth]{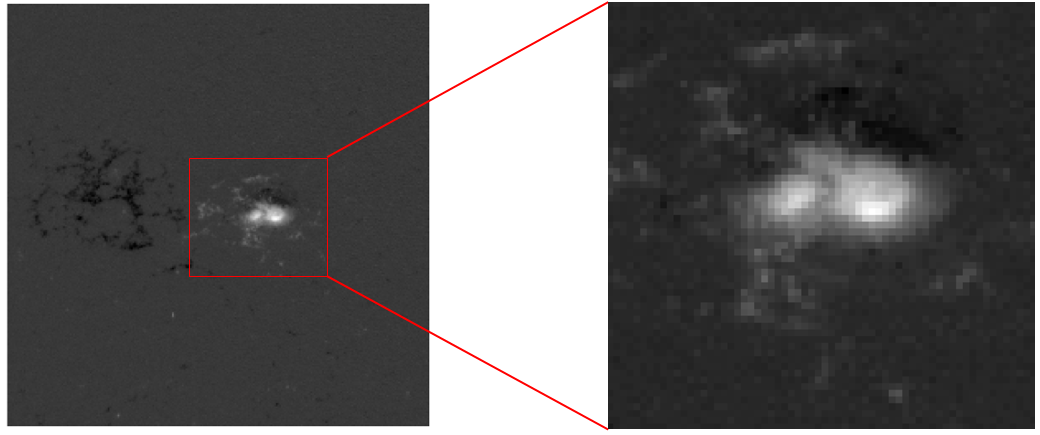}%
}\hspace{0.05\linewidth}
\subfloat[LDM residual\label{fig:ldmres_zoom_crop}]{%
 \includegraphics[width=0.58\linewidth]{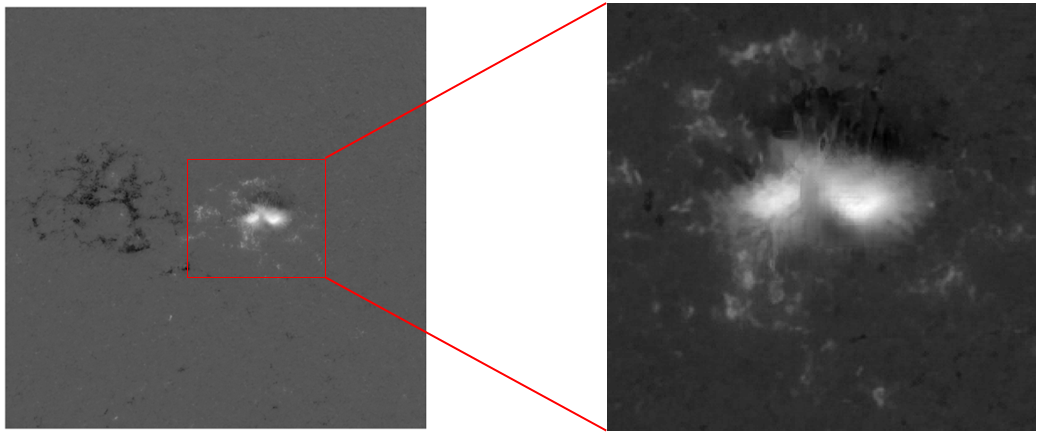}%
}\hspace{0.05\linewidth}
\subfloat[Progressive\label{fig:prog_ex}]{%
 \includegraphics[width=0.58\linewidth]{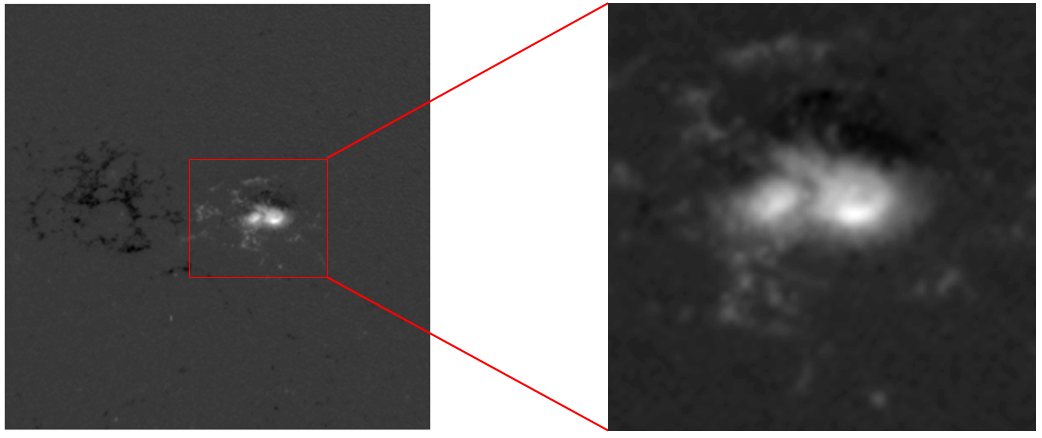}%
}\hspace{0.05\linewidth}
\subfloat[Enhance\label{fig:enha_ex}]{%
 \includegraphics[width=0.58\linewidth]{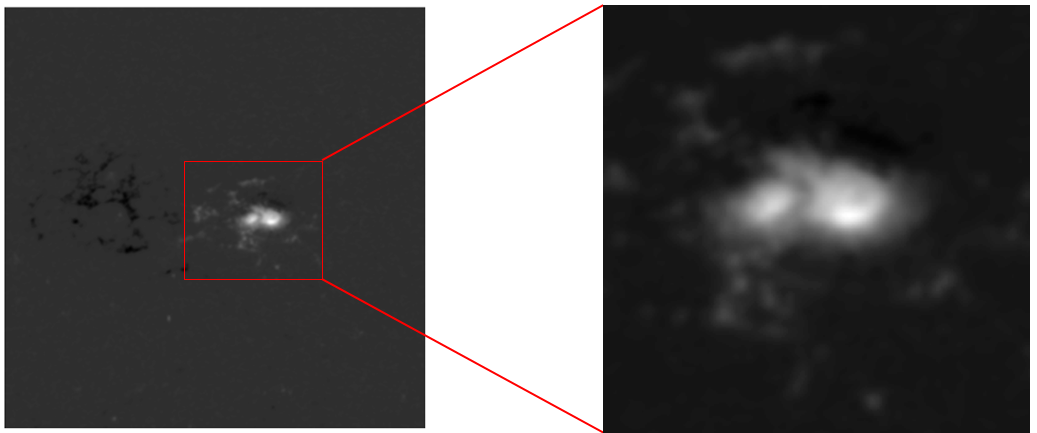}%
}
\caption{Prediction example of the MDI to HMI application. NOOA AR number 11108 from 22 September 2010.}
\label{fig:mdi_hmi_orediction}
\end{figure*}

The results of this experiment are shown in Table \ref{tab:results_metrics}.
We take 250 images from the validation set and use SunPy \citep{sunpy_community2020} to identify the most intense AR present, as described in Section \ref{sec:metrics}. A 1024$\times$1024 pixel crop is taken around the AR, downscaled to 256$\times$256 pixels, and then, depending on the model used (diffusion or not), upscaled back to 1024$\times$1024 pixels as previously mentioned (Figure \ref{fig:downscaled}). Furthermore, we super-resolve the 250 test images 10 times each using the diffusion model frameworks. Since these models are probabilistic, we can assess the stability of their predictions given a certain input image. This is not possible with the Enhance and Progressive approaches, as they are deterministic, meaning that each prediction is always the same given the same input data.

As shown in Table \ref{tab:results_metrics}, the best-performing model in terms of PSNR and SSIM is the Progressive model. The Progressive model \citep{Rahman2020} is a deterministic model trained using MSE loss in the pixel space. As explained in Section \ref{sec:sr}, the Super-Resolution problem is highly ill-posed, meaning that for each pair of LR and HR images, there are an infinite number of possible predicted HR images. Training a model with MSE loss in the pixel space results in an output that averages over all possible outcomes, leading to blurry predictions, as observed in Figure \ref{fig:enhance} and Figure \ref{fig:progressive}. This blurriness occurs because the model is uncertain about how to super-resolve fine-grained details. As explained in Section \ref{sec:metrics}, both PSNR and SSIM tend to favor models that produce blurry outputs, which explains why the Progressive model excels in terms of PSNR, and why the Enhance model also performs well in terms of SSIM.

\begin{figure}
\centering
 \includegraphics[width=\hsize]{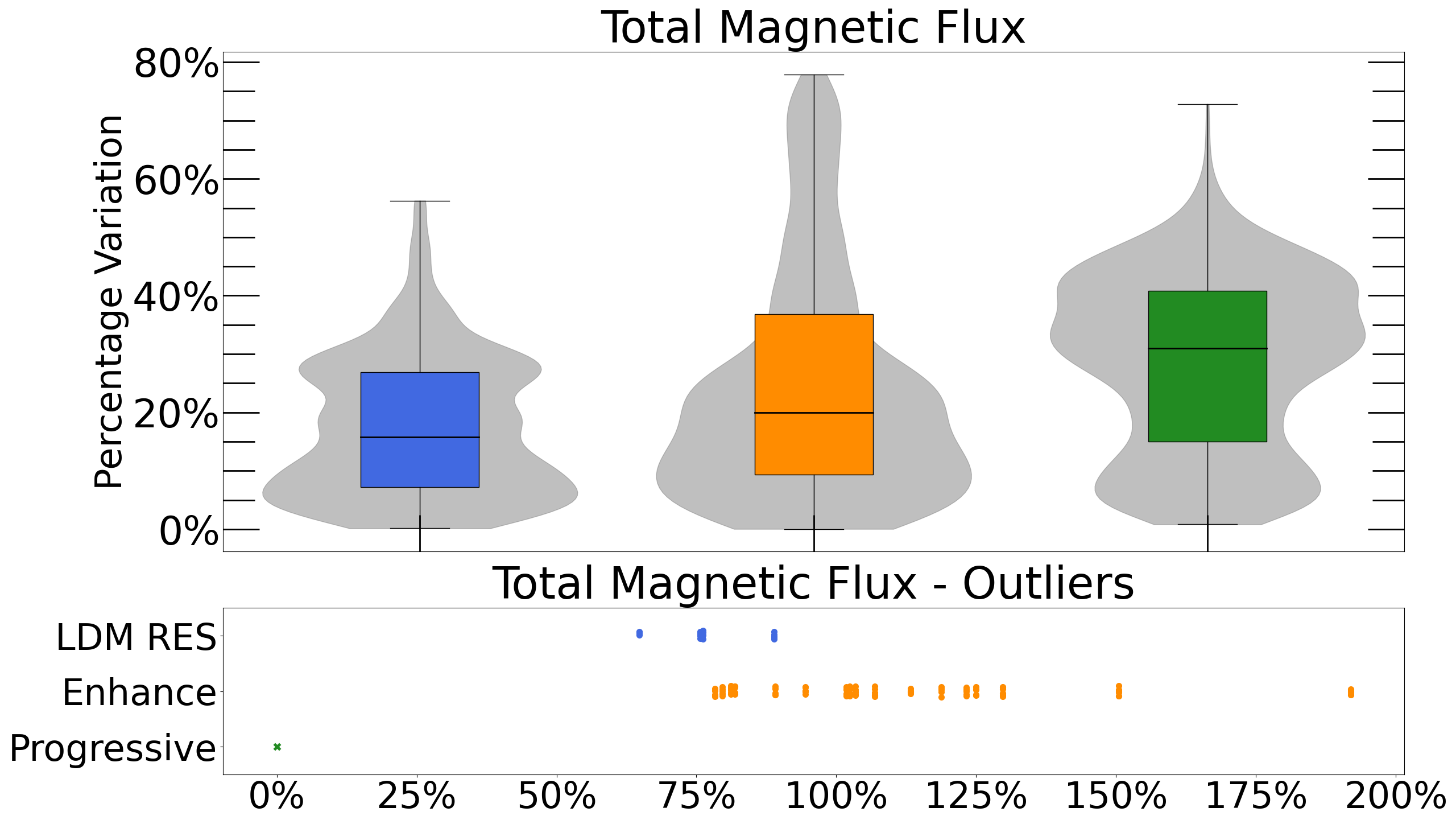}%
\caption{The image shows the boxplot of the unsigned magnetic flux metric, whose mean and standard deviation are presented in Table \ref{tab:results_mdi2hmi}. The main plot displays the data distribution between the 25th and 75th percentiles, while the subplot highlights all outliers outside this range. The boxplot is included because, although the mean values in Table \ref{tab:results_mdi2hmi} are acceptable, the wide confidence intervals make it easier to visually assess the model's quality.}
\label{fig:unsigned_mag_flux}
\end{figure}

\begin{figure}
\centering
 \includegraphics[width=\hsize]{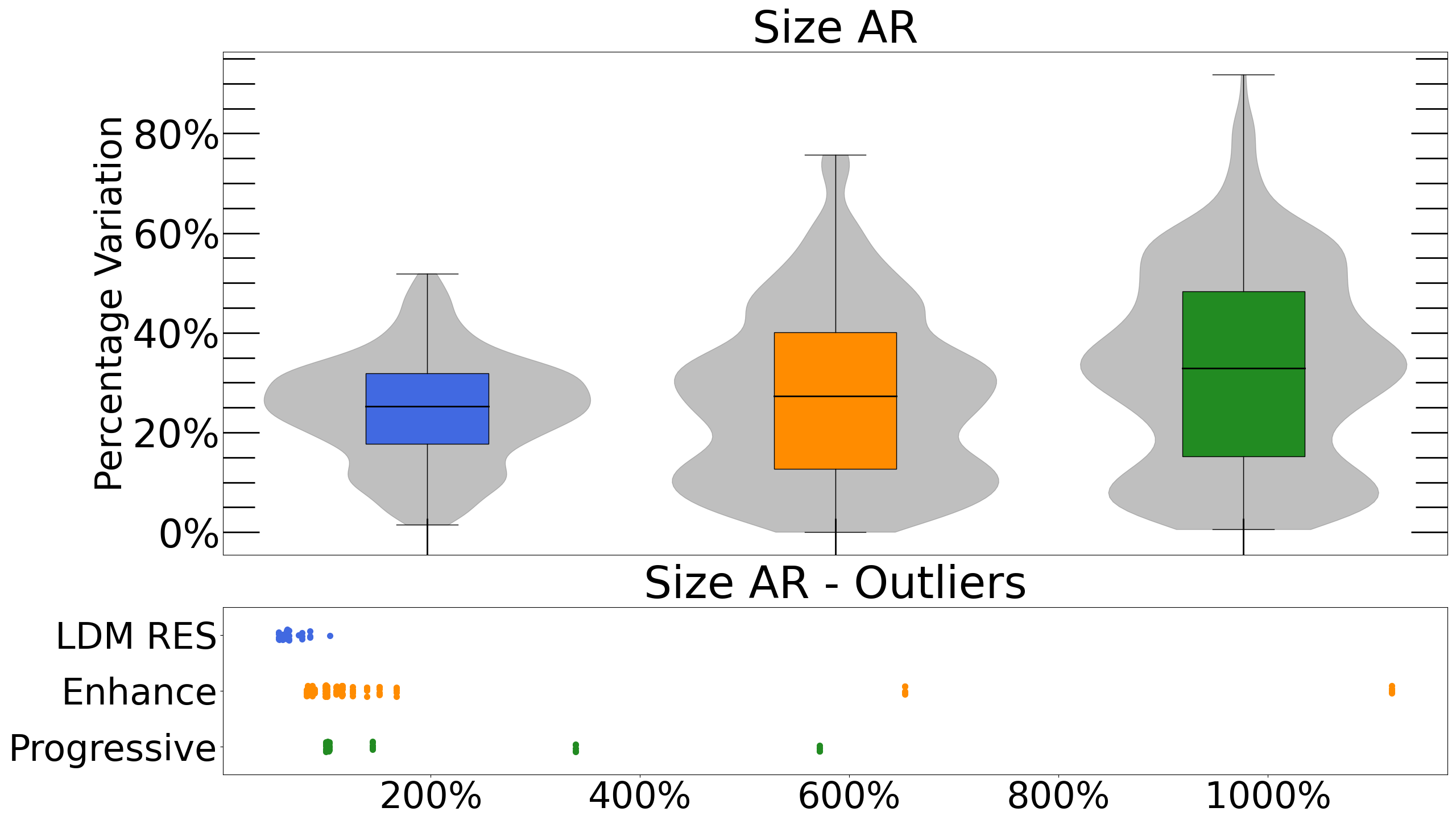}%
\caption{The image shows the boxplot of the active region size metric, whose mean and standard deviation are presented in Table \ref{tab:results_mdi2hmi}. The main plot displays the data distribution between the 25th and 75th percentiles, while the subplot highlights all outliers outside this range. The boxplot is included because, although the mean values in Table \ref{tab:results_mdi2hmi} are acceptable, the wide confidence intervals make it easier to visually assess the model's quality.}
\label{fig:size_ar_boxplot}
\end{figure}

Following the approach in \cite{Saharia2021}, we aim to test metrics that better align with human perception. Therefore, we use both the FID and LPIPS distances. LPIPS  is a pairwise metric that evaluates the perceptual similarity between two images. It focuses on how similar a generated image is to a reference image based on deep features, capturing fine details of human visual perception. On the other hand, the FID assesses both the quality and diversity of generated images by comparing the distribution of generated images to that of original images. Unlike LPIPS, FID captures the overall dataset spread, providing a broader view of how well the generated samples align with the real data distribution regarding perceptual features.
The LDM with residual is the best model in terms of FID and LPIPS as shown in Table \ref{tab:results_metrics}. This result is also consistent with what can be inferred by a visual inspection of Figure \ref{fig:visual_inspections}. Nevertheless, we are interested in more than just the visual quality of the image. Specifically, we want to assess whether the model can preserve the underlying physics while improving the image appearance. Consequently, we use the unsigned magnetic flux and the size of ARs as metrics. The unsigned magnetic flux is crucial because its variations are correlated with the structural complexity of the magnetic field, which is essential for understanding and predicting energetic events such as flares or coronal mass ejections (CMEs) \citep{Wiegelmann2014}. It is important that the model enhances the visual appeal of the images without introducing artefacts that alter the unsigned magnetic flux. The size of the AR, which we compute by counting the number of pixels inside the contour with the OpenCV library \citep{opencv_library}, as described in \ref{sec:metrics}, is an important indicator of potential flares or other eruptive events \citep{Toriumi2019}.

The unsigned magnetic flux and the AR size results in Table \ref{tab:results_metrics} are presented in forms of percentage variation with respect to the same value computed on the target magnetograms. To compute those metrics we go back from the predicted values of network between -1 and 1 to the original range -3000 G and +3000 G by reverting the normalisation used. We measure the pixel area in meters and then sum the pixels to obtain the magnetic field value multiplied by the underlying area. For both the unsigned magnetic flux and the size of the AR, the LDM with residuals performs best. This suggests we are not introducing significant artefacts that alter the magnetic flux or AR size. Thanks to the finer details, we can estimate these values more accurately compared to the Progressive and Enhance models, whose results are affected by the blurriness that can be observed in Figure \ref{fig:visual_inspections}.

Based on the discussion and results, the LDM with residuals is the best overall performer. It excels in preserving the underlying physical properties, such as unsigned magnetic flux and AR size, while producing high-quality images with minimal artefacts. Despite the Progressive model's strong performance in metrics like PSNR and SSIM, which favour smoother but blurrier outputs, the LDM with residuals outperforms perceptually important metrics such as FID and LPIPS. These metrics better capture fine details and human visual perception, making the LDM with residuals a more suitable model for Super-Resolution tasks where visual quality and physical accuracy are critical. Its ability to better estimate values like the magnetic flux and AR size without introducing significant artefacts confirms its superiority for this specific application.

\section{Super-Resolution of MDI magnetograms}
\label{sec:mdi_app}
After finding the best model for our $\times$4 super resolution task in a controlled environment, where the only difference among the LR and the HR images is the spatial resolution, we apply this model to a real case scenario to upscale the MDI/SOHO LoS magnetograms. This approach allows us to use more data and extend the training beyond the limited period from May 1, 2010, to April 11, 2011. By creating synthetic low-resolution/high-resolution pairs from HMI over a larger time span, we expose the model to a wider variety of solar features and teach it to super-resolve these effectively. This is crucial for applying the model to data from 1995 to 2010, for which we do not have an HMI counterpart. The input MDI data are normalised as the HMI data between -1 and 1 but before doing that we scale the MDI data by 1/1.4 following the paper by \cite{Liu2012}.

MDI/SOHO has a spatial resolution of 2"/pixel, while HMI/SDO operates at a higher resolution of 0.5"/pixel. The only period during which MDI and HMI were both operational was from May 1, 2010, to April 11, 2011, as outlined in Section \ref{sec:data_source}. Therefore, we collect paired MDI/HMI data from this period, using HMI as the ground truth for comparison, with HMI serving as our HR reference data in this context. We can see an example of an AR seen by both the instruments in Figure \ref{fig:mdi_hmi}. 

\begin{figure*}
\centering
 \includegraphics[width=0.9\hsize]{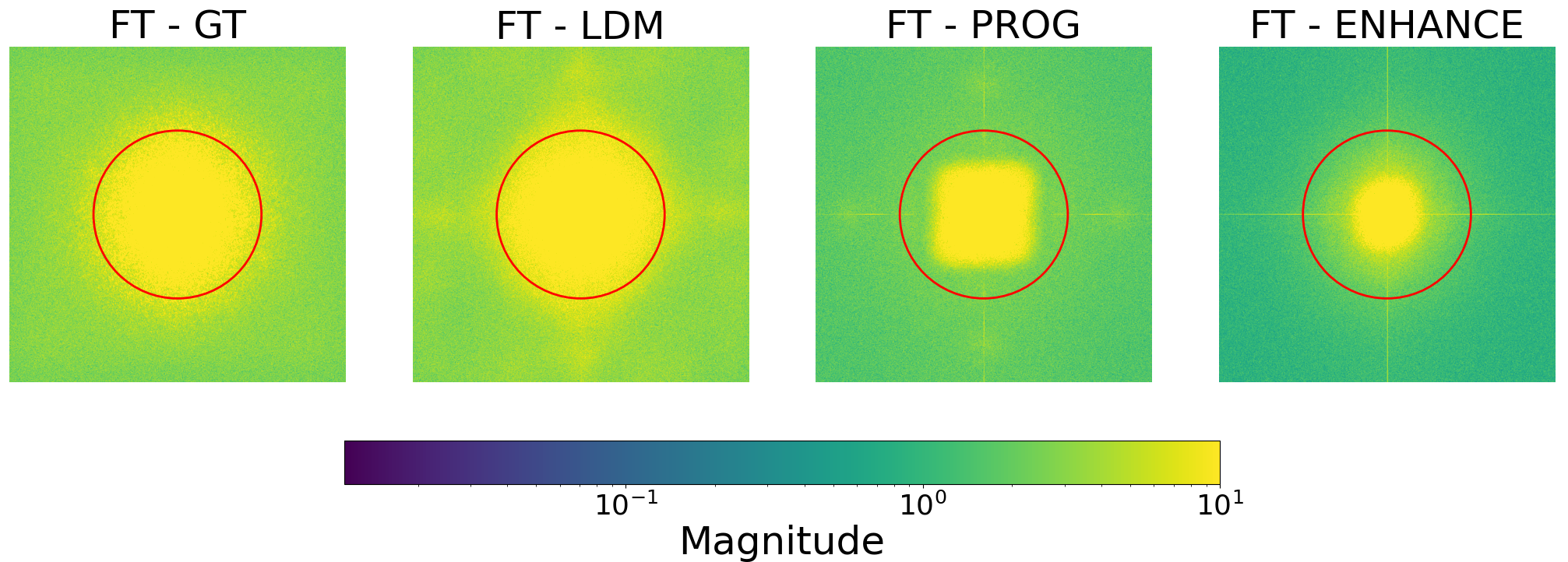}%
\caption{The images represent the amplitude of the Fourier Transform of the Ground Truth image (FT - GT), and of the predictions obtained with our Latent Diffusion Model (FT - LDM), with the Progressive Model (FT - PROG), and with the Enhance Model (FT - ENHANCE). The red circles indicate the Fourier frequencies corresponding to a spatial resolution of 2''. This visualization highlights how each model predicts or blurs high-frequency features, which are crucial for capturing fine details.}
\label{fig:ft_circle}
\end{figure*}

We finetune our pre-trained residual LDM on pairs of actual MDI/HMI data, allowing the model to first learn the Super-Resolution task and then calibrate between the two instruments, as discussed in \citep{Jaramillo2024}. In this setup, the inputs to our model are MDI data. MDI images have a size of 1024 $\times$ 1024 pixels, so we identify the most intense active regions (ARs) using SunPy \citep{sunpy_community2020}, take a 256 $\times$ 256 pixels crop around them, and then upscale the image by replicating each pixel 4 times, as described in Section \ref{sec:experiments}.

To finetune our pre-trained residual LDM we use the Low-Rank Adaption (LoRA) technique \cite{hu2021loralowrankadaptationlarge} for the self attention layers and we unfreeze the bottleneck layers of the U-Net. LoRA is used to finetune the self-attention layers in a more computationally efficient way. By decomposing the weight matrices into low-rank updates, LoRA allows the model to adapt effectively without retraining all parameters, saving time and reducing the risk of overfitting. The bottleneck layers in the U-Net architecture are critical for capturing the compressed representation of the image. By unfreezing these layers during fine-tuning, the model can better adapt to the specific characteristics of the MDI/HMI data and learn more detailed representations specific to the new MDI instrument. We finetune the residual LDM with 200 MDI/HMI data from May 1, 2010 to August 31, 2010 and test it on 200 random pairs from September 1, 2010 to April 11, 2011.

We focus our testing on the LDM with residuals, as it is the best-performing diffusion framework among the four tested. However, we also compare its performance against the two deterministic approaches, Enhance and Progressive. Both Enhance and Progressive models are fine-tuned on the same 200 MDI/HMI pairs as the LDM and are tested on the same 200 pairs for consistency.

As shown in Table \ref{tab:results_mdi2hmi}, we observe a similar trend to what is seen in Table \ref{tab:results_metrics} for the LR HMI to HR HMI images. The LDM with residuals performs best in terms of LPIPS, FID, and the physics-based metrics. However, for pixel-level metrics like PSNR and SSIM, the Enhance model outperforms the others. The reason for this, as discussed in Section \ref{sec:experiments}, is that the outputs of the Progressive and Enhance models tend to be blurrier, as also evident in Figure \ref{fig:mdi_hmi_orediction}. In addition, we provide the boxplots of the unsigned magnetic flux (Figure \ref{fig:unsigned_mag_flux}) and the active region size (Figure \ref{fig:size_ar_boxplot}) metrics to help the reader better understand the model quality presented in Table \ref{tab:results_mdi2hmi}, since that although the mean values are acceptable, the confidence intervals are relatively wide.

The goal of this Super-Resolution task is to apply it to all MDI data from 1995 to 2010, for which we do not have corresponding HMI data, and enhance the spatial resolution. This will allow us to overcome the resolution limitations of the MDI instrument and study past events at a higher resolution, comparable to modern instruments. To ensure we can actually extract relevant information through 4x Super-Resolution from the original 2"/pixel data, we test the results in the Fourier domain. Specifically, we check for the presence of high-frequency signals below 2" in the generated images and assess the model confidence in these predictions.

We compute for each model prediction the 2D Fast Fourier Transform (FFT) to convert the spatial domain images into the frequency domain. The zero frequency component is then shifted to the centre for easier interpretation. Afterwards, the magnitudes of the shifted FFT results are calculated, allowing us to evaluate and visualise the frequency content of each image. The red circles indicate frequencies corresponding to 2'' resolution in the physical space. Specifically, high values of the FFT amplitude inside the circles indicate presence of spatial features lager than 2'' in the predicted magnetograms, while large values outside the circles indicate presence of features smaller than 2'' in the magnetograms. We show in Figure \ref{fig:ft_circle} the amplitude of the Fourier Transform of the ground truth image and of the magnetograms predicted with our LDM, the Progressive model, and the Enhance model. We saturate the pixel values above a specific threshold to better visualize the intensities of the high frequency Fourier components, which lie outside the circle. We observe that both the Progressive and Enhance models fail to predict the high-frequency details, in line with the discussion in Section \ref{sec:results}. This is because these models average over all possible solutions that map a low-resolution image to a high-resolution one, resulting in blurred outputs without significant high-frequency content. In contrast, the LDM with residuals retains high-frequency details, as the pixel intensities closely resemble those of the ground truth. This observation is consistent with the results in Tables \ref{tab:results_metrics} and \ref{tab:results_mdi2hmi}, where the FID and LPIPS metrics favour the LDM with residuals. These metrics, which align well with human visual perception, effectively capture the presence or absence of fine details in the images.

\begin{figure}
\centering
 \includegraphics[width=\hsize]{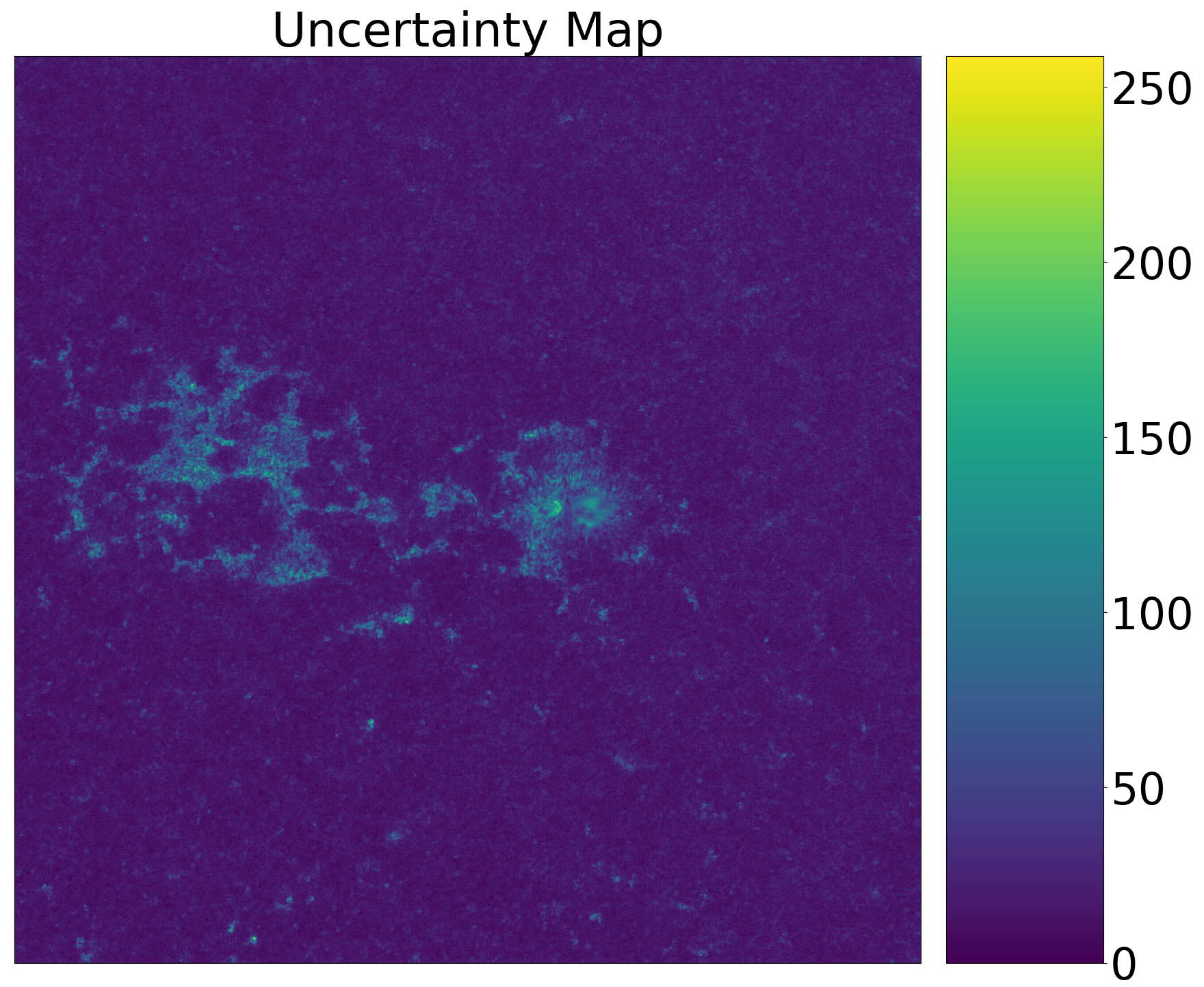}%
\caption{The image shows a standard deviation map derived from 10 repeated  model predictions. Input image September 22, 2010 on AR 11108. The standard deviation values are expressed in G.}
\label{fig:std_map}
\end{figure}

\begin{figure*}
\centering
 \includegraphics[width=\hsize]{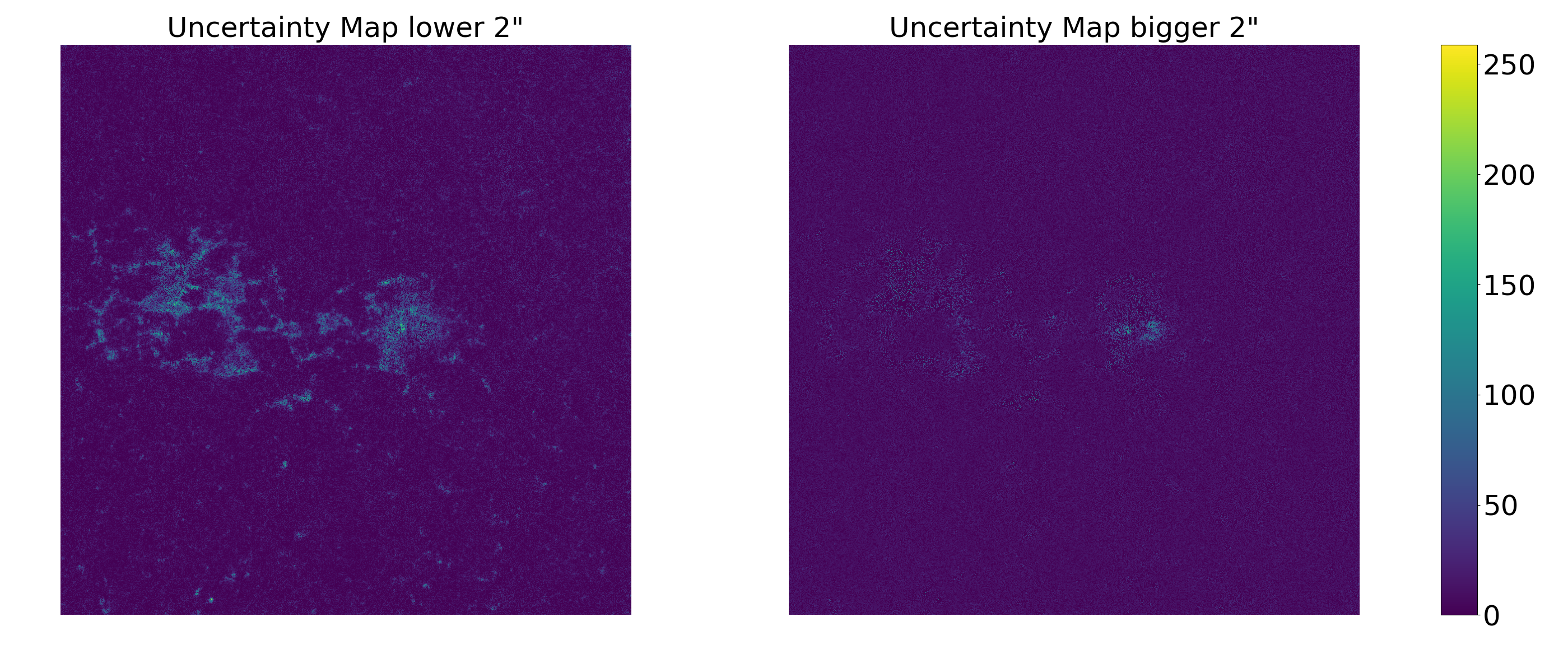}%
\caption{Comparison of the uncertainty maps of the features lower than 2" and larger than 2" obtained with the application of the Butterworth high-pass filter \citep{butterworth1930filter} to the uncertainty map in Figure \ref{fig:std_map}. The values on the colorbar are expressed in G.}
\label{fig:std_filtered}
\end{figure*}

However, we aim not only to verify if these high-frequency features (smaller than 2"), which are absent in the input MDI map, are predicted but also to assess their associated uncertainties. This analysis is not feasible with classical deterministic models. On the contrary with the LDMs we can perform this evaluation due to their stochastic nature, as it is done in \cite{Ramunno2024mag2mag}. To visualise the uncertainties of the LDM model, we super-resolve the same MDI magnetogram ten times, using AR 11108 from September 22, 2010 as an example. We then concatenate the ten predictions along the channel dimension and compute the standard deviation for each pixel. The resulting image shows pixel-wise uncertainty, where higher values indicate greater standard deviation, meaning the model is less confident in its prediction for that pixel. This allows us to associate a specific uncertainty measure with each pixel (Figure \ref{fig:std_map}).

Additionally, while we aim to predict features smaller than 2", it is important that the model exhibits greater uncertainty in these predictions compared to the larger features. This is because we do not want the model to modify the existing features present in the MDI map. To assess this, we isolate the uncertainties between features larger and smaller than 2" by applying a Butterworth high-pass filter \citep{butterworth1930filter} to the uncertainty map, which allows us to separate the high-frequency components (representing features smaller than 2"). We then, thanks to the filtered uncertainty map, create a mask and overlay it to the uncertainty map on Figure \ref{fig:std_map}.

We observe in Figure \ref{fig:std_filtered} that uncertainties are higher for features smaller than 2" compared to features larger than 2". Although uncertainties are also present in the right Figure \ref{fig:std_filtered}, they correspond to areas of high-intensity pixels (Figure \ref{fig:mdi_hmi}), which results in a relatively lower uncertainty.

\section{Conclusions}
\label{sec:conclusions}

In this work, we present a novel method based on Latent Diffusion Models to achieve Super-Resolution for solar magnetograms, specifically focusing on the data from the MDI and the HMI instruments. Our approach successfully addresses the challenge of enhancing the spatial resolution of MDI data from 2"/pixel to match the 0.5"/pixel resolution of HMI. By leveraging a pre-trained autoencoder to reduce the image size and applying residual learning, we demonstrated significant improvements in both the visual quality of super resolved images and the preservation of underlying physical properties, such as the unsigned magnetic flux and the ARs.

Our experiments (Table \ref{tab:results_metrics}) showed that the LDM with residuals outperforms deterministic models, such as the Enhance and Progressive models, in terms of perceptual metrics like LPIPS and FID, which are crucial for assessing the fine details in high-resolution images. Moreover, the LDM with residual is easily generalizable to other instruments as we can see in Table \ref{tab:results_mdi2hmi} where we finetune on a small amount of MDI/HMI pairs to apply it for super resolving MDI magnetograms. 

Most importantly, the LDM with residuals can generate an uncertainty map due to its stochastic behaviour, allowing us to identify where the model struggles the most (Figure \ref{fig:std_map}). Using the Fourier transform (Figure \ref{fig:ft_circle}) and the uncertainty maps, we demonstrate that our model can super-resolve features smaller than 2" while also assessing their reliability. This process is essential because, if we cannot predict features smaller than 2", we are merely enhancing the image aesthetics without adding meaningful information. As shown in Figure \ref{fig:ft_circle}, deterministic models fail in this regard.

Furthermore, we demonstrate that the LDM with residuals does not sacrifice features larger than 2" in order to predict smaller ones (Figure \ref{fig:std_filtered}). This is crucial because it ensures that we are preserving the existing knowledge in the data. Although the model shows higher uncertainty for smaller features, this is not problematic. On the contrary, it gives a hint on the presence of an hidden feature, especially when super resolving pre-2010 data where we lack HMI counterparts. This allows us to study features that were previously invisible.

However, we acknowledge the potential bias introduced by fine-tuning on a limited dataset that represents only a specific phase of the solar cycle (May 2010 – April 2011), predominantly capturing the rising phase of solar activity. This limited period restricts the diversity of solar conditions for fine-tuning. Since super-resolution models rely on learned priors from training data rather than exact recovery of missing information, biases can arise. To mitigate this, we employed multiple strategies, including perceptual metrics like FID, uncertainty maps to assess prediction reliability, and Fourier domain analysis to ensure the presence of features smaller than 2” without introducing artificial structures. Additionally, we tested the model on unseen MDI data from different periods to confirm robustness beyond the fine-tuning set. Despite these efforts, some bias related to the solar cycle phase may remain, which should be considered when interpreting results.

Applying this technique to MDI data from Solar Cycle 23 opens up exciting possibilities for reanalysing past solar events with a resolution comparable to modern instruments like HMI. Furthermore, it is also helpful from a generative point of view because we can generate images in a smaller pixel resolution \citep{Ramunno2024mag2mag} and then super resolve them. Future work will aim to extend this approach to enhance the temporal resolution of MDI, offering a more detailed view of the dynamic evolution of solar features.

\section{Data Availability}
Visit this \href{https://github.com/fpramunno/ldm_superresolution}{https://github.com/fpramunno/ldm\_superresolution} for the code.
\begin{acknowledgements}
      This research was partially funded by the SNF Sinergia project (CRSII5-193716): Robust Deep Density Models for High-Energy Particle Physics and Solar Flare Analysis (RODEM). The authors would like to thank Manolis K. Georgoulis from Johns Hopkins APL for their valuable discussions and insightful feedback that helped shape this work.
\end{acknowledgements}

\bibliographystyle{aa} 
\bibliography{references} 

\newpage
\begin{appendix} 

\section{Computational Time Comparison}
\label{sec:comp_time}

We trained our latent-space diffusion model for 30 epochs on an NVIDIA TITAN X GPU, with a total training time of 6 days, 20 hours, 17 minutes, and 26 seconds. For inference, the model takes approximately 40 seconds to super-resolve a single image using 1,000 timesteps on the same hardware.
To provide context, we also trained the same diffusion model directly in pixel space, which required around 8 days for 30 epochs. While the reduction in training time is relatively modest (about 14.6\%), the main advantage of operating in latent space lies in the significantly improved inference speed. Latent-space inference is approximately 2.5 times faster, reducing the super-resolution time for a single image from 1 minute and 41 seconds to just 40 seconds. This highlights the computational efficiency and practicality of using latent-space diffusion models, particularly for large-scale inference tasks.

\end{appendix} 

\end{document}